\documentclass[%
 aip,
 amsmath,amssymb,
]{revtex4-1}
\usepackage{scalerel}
\usepackage{tikz}
\usetikzlibrary{svg.path,calc,shapes,arrows,positioning,shadows,shadings}
\usepackage[colorlinks=false,linkcolor=blue]{hyperref}
\usepackage{graphicx}
\usepackage{dcolumn}
\usepackage{bm}
\usepackage[utf8]{inputenc}
\usepackage[T1]{fontenc}
\usepackage{mathptmx}
\usepackage[ruled,vlined]{algorithm2e}
\usepackage{etoolbox}
\usepackage{chemfig}
\usepackage{braket}
\usepackage{simpler-wick}
\usepackage{bm}
\usepackage{float}
\usepackage[version=4]{mhchem}
\usepackage{booktabs}
\usepackage{calc}
\usepackage{booktabs}       
\usepackage{threeparttable}  
\usepackage{adjustbox}  
\usepackage{rotating}

\usepackage{siunitx}
\usepackage{booktabs}
\usepackage{pgfplots}
\makeatletter
\def\@email#1#2{%
 \endgroup
 \patchcmd{\titleblock@produce}
  {\frontmatter@RRAPformat}
  {\frontmatter@RRAPformat{\produce@RRAP{*#1\href{mailto:#2}{#2}}}\frontmatter@RRAPformat}
  {}{}
}%
\makeatother
\usetikzlibrary{calc}
\usetikzlibrary{decorations.pathreplacing,decorations.markings}

\tikzset{
  on each segment/.style={
    decorate,
    decoration={
      show path construction,
      moveto code={},
      lineto code={
        \path [#1]
        (\tikzinputsegmentfirst) -- (\tikzinputsegmentlast);
      },
      curveto code={
        \path [#1] (\tikzinputsegmentfirst)
        .. controls
        (\tikzinputsegmentsupporta) and (\tikzinputsegmentsupportb)
        ..
        (\tikzinputsegmentlast);
      },
      closepath code={
        \path [#1]
        (\tikzinputsegmentfirst) -- (\tikzinputsegmentlast);
      },
    },
  },
  mid arrow/.style={postaction={decorate,decoration={
        markings,
        mark=at position .62 with {\arrow[#1]{>}}
      }}},
  mid rarrow/.style={postaction={decorate,decoration={
        markings,
        mark=at position .38 with {\arrow[#1]{<}}
      }}},
  mid darrow/.style={postaction={decorate,decoration={
        markings,
        mark=at position .62 with {\arrow[#1]{>>}}
      }}},
  mid rdarrow/.style={postaction={decorate,decoration={
        markings,
        mark=at position .38 with {\arrow[#1]{<<}}
      }}},
}

\pgfdeclaredecoration{complete sines}{initial}
{
  \state{initial}[
        width=+0pt,
        next state=sine,
        persistent precomputation={\pgfmathsetmacro\matchinglength{
            \pgfdecoratedinputsegmentlength / int(\pgfdecoratedinputsegmentlength/\pgfdecorationsegmentlength)}
            \setlength{\pgfdecorationsegmentlength}{\matchinglength pt}
        }] {}
  \state{sine}[width=\pgfdecorationsegmentlength]{
        \pgfpathsine{\pgfpoint{0.25\pgfdecorationsegmentlength}{0.5\pgfdecorationsegmentamplitude}}
        \pgfpathcosine{\pgfpoint{0.25\pgfdecorationsegmentlength}{-0.5\pgfdecorationsegmentamplitude}}
        \pgfpathsine{\pgfpoint{0.25\pgfdecorationsegmentlength}{-0.5\pgfdecorationsegmentamplitude}}
        \pgfpathcosine{\pgfpoint{0.25\pgfdecorationsegmentlength}{0.5\pgfdecorationsegmentamplitude}}
  }
  \state{final}{}
}

\edef\xcoord{11} \edef\xcoor{11} \edef\xcoorbeg{11} \edef\xcoorend{11} \edef\result{11} \edef\ycoor{11} \edef\ycoord{11}
\edef\heffsize{6}
\pgfmathsetmacro{\scalh}{1} 
\pgfmathsetmacro{\scalv}{1} 
\pgfmathsetmacro{\pixtocoord}{(1/25)} 
\pgfmathsetmacro{\xcoord}{(-1*\scalh)} \pgfmathsetmacro{\ycoord}{(-1*\scalv)} 
\pgfmathsetmacro{\xcoor}{\xcoord} \pgfmathsetmacro{\ycoor}{(\ycoord+1*\scalv)} 
\pgfmathsetmacro{\xcoordg}{\xcoord}
\pgfmathsetmacro{\xcoorH}{\xcoord}
\pgfmathsetmacro{\yvac}{(\ycoor+0.5*\scalv)} \pgfmathsetmacro{\xvac}{(0.25*\scalh)} \pgfmathsetmacro{\xdvac}{(0.5*\scalh)}
\newcounter{stoer}

\tikzset{ph-line-arrow-save/.style={>=stealth,thin,postaction={on each segment={mid arrow}}}}
\tikzset{ph-liner-arrow-save/.style={>=stealth,thin,postaction={on each segment={mid rarrow}}}}
\tikzset{ph-line-darrow-save/.style={>=stealth,thin,postaction={on each segment={mid darrow}}}}
\tikzset{ph-liner-darrow-save/.style={>=stealth,thin,postaction={on each segment={mid rdarrow}}}}
\tikzset{ph-line-noarrow-save/.style={thin}}
\tikzset{ph-line-save/.style={ph-line-arrow-save}}
\tikzset{ph-liner-save/.style={ph-liner-arrow-save}}
\tikzset{exoper-line-save/.style={very thick, solid}}
\tikzset{exvac-line-save/.style={thick, dotted}}
\tikzset{hoper-line-fey-save/.style={ decoration={complete sines,segment length=0.1cm,amplitude=0.1cm}, decorate}}
\tikzset{hoper-line-save/.style={very thick, dashed}}
\tikzset{hoperbar-line-save/.style={ hoper-line-save, double}}
\tikzset{heffoper-line-save/.style={thick, solid, double distance = \heffsize, line cap=rect }}
\tikzset{exoper2-line-save/.style={ thick, solid, double}}
\tikzset{exoper3-line-save/.style={ thick, transparent}}
\tikzset{exoper-line/.style={exoper-line-save}}
\tikzset{hoper-line/.style={hoper-line-save}}
\tikzset{heffoper-line/.style={heffoper-line-save}}
\tikzset{ph-line/.style={ph-line-save}}

\newcommand{\bdiag}[1][]{
 \ifx&#1&%
  \begin{tikzpicture}
 \else
  \begin{tikzpicture}[scale=#1]
  \pgfmathsetmacro{\pixtocoord}{\pixtocoord/#1}
 \fi
}

\newcommand{\ediag}{\end{tikzpicture}}

\newcommand{\dorigin}[1][]{
 \ifx&#1&%
  \pgfmathsetmacro{\xorig}{0}
 \else
  \pgfmathsetmacro{\xorig}{#1}
 \fi
 \node(Orig) at (\xorig,0){};
}

\newcommand{\dAmp}[4][]{
\pgfmathsetmacro{\xcoorbeg}{(\xcoor+0.5*\scalh)} 
\pgfmathsetmacro{\xcoorend}{(\xcoorbeg+#3*\scalh/2)} 
\draw[exoper-line](\xcoorbeg,\ycoord) -- (\xcoorend,\ycoord);
\pgfmathsetmacro{\result}{(\xcoorend+0.35*\scalh)} 
\ifx&#1&%
  \pgfmathsetmacro{\xcoor}{\xcoorend}
\else
  \node at (\result,\ycoord){#1};
  \pgfmathsetmacro{\xcoor}{(\xcoorend+0.25*\scalh)}
\fi
\ifnum#3=1
 \pgfmathsetmacro{\xxx}{((\xcoorend-\xcoorbeg)/2)} 
 \pgfmathsetmacro{\result}{\xcoorbeg}
\else
 \pgfmathsetmacro{\xxx}{((\xcoorend-\xcoorbeg)/( #3 - 1 ))} 
 \pgfmathsetmacro{\result}{(\xcoorbeg-\xxx)}
\fi
\foreach \x in {1,...,#3}
{
  \coordinate (vertex) at ($(\result,\ycoord)+(\x*\xxx,0)$);
  \node[inner sep=0pt,minimum size=0pt] (#4\x) at  (vertex) {};
  \coordinate (vtxvac) at ($(\result,\yvac)+(\x*\xxx,0)$);
  \node[inner sep=0pt,minimum size=0pt] (#4\x v1) at ($(vtxvac)-(\xvac,0)$) {};
  \node[inner sep=0pt,minimum size=0pt] (#4\x v2) at ($(vtxvac)+(\xvac,0)$) {};
}
\node[inner sep=0pt,minimum size=0pt] (#4) at  (#41) {};
\node[inner sep=0pt,minimum size=0pt] (#4v1) at (#41v1) {};
\node[inner sep=0pt,minimum size=0pt] (#4v2) at (#41v2) {};
\ifx&#2&%
\else  
\pgfmathsetmacro{\result}{((\xcoorend+\xcoorbeg)/2)} 
\setcounter{stoer}{#2} 
\node[inner sep=0pt,minimum size=0pt] at (\result,\ycoord){{\footnotesize\slshape\sffamily \Roman{stoer}}}; 
\fi
}

\newcommand{\dAmpD}[4][]{
\pgfmathsetmacro{\xcoorbeg}{(\xcoordg+0.5*\scalh)} 
\pgfmathsetmacro{\xcoorend}{(\xcoorbeg+#3*\scalh/2)} 
\draw[exoper-line](\xcoorbeg,\yvac) -- (\xcoorend,\yvac);
\pgfmathsetmacro{\result}{(\xcoorend+0.35*\scalh)} 
\ifx&#1&%
\pgfmathsetmacro{\xcoordg}{\xcoorend}
\else
\node at (\result,\yvac){#1};
\pgfmathsetmacro{\xcoordg}{(\xcoorend+0.25*\scalh)}
\fi
\ifnum#3=1
 \pgfmathsetmacro{\xxx}{((\xcoorend-\xcoorbeg)/2)} 
 \pgfmathsetmacro{\result}{\xcoorbeg}
\else
 \pgfmathsetmacro{\xxx}{((\xcoorend-\xcoorbeg)/( #3 - 1 ))} 
 \pgfmathsetmacro{\result}{(\xcoorbeg-\xxx)}
\fi
\foreach \x in {1,...,#3}
{
  \coordinate (vertex) at ($(\result,\yvac)+(\x*\xxx,0)$);
  \node[inner sep=0pt,minimum size=0pt] (#4\x) at  (vertex) {};
  \coordinate (vtxvac) at ($(\result,\ycoord)+(\x*\xxx,0)$);
  \node[inner sep=0pt,minimum size=0pt] (#4\x v1) at ($(vtxvac)-(\xdvac,0)$) {};
  \node[inner sep=0pt,minimum size=0pt] (#4\x v2) at ($(vtxvac)+(\xdvac,0)$) {};
}
\node[inner sep=0pt,minimum size=0pt] (#4) at  (#41) {};
\node[inner sep=0pt,minimum size=0pt] (#4v1) at (#41v1) {};
\node[inner sep=0pt,minimum size=0pt] (#4v2) at (#41v2) {};
\ifx&#2&%
\else
\pgfmathsetmacro{\result}{((\xcoorend+\xcoorbeg)/2)} 
\setcounter{stoer}{#2} 
\node[inner sep=0pt,minimum size=0pt] at (\result,\yvac){{\footnotesize\slshape\sffamily \Roman{stoer}}}; 
\fi
}

\newcommand{\dHeff}[4][]{
\pgfmathsetmacro{\xcoorbeg}{(\xcoorH+0.5*\scalh)} 
\pgfmathsetmacro{\xcoorend}{(\xcoorbeg+#3*\scalh/2)} 
\draw[heffoper-line](\xcoorbeg,\ycoor) -- (\xcoorend,\ycoor);
\pgfmathsetmacro{\ycoora}{(\ycoor+\pixtocoord*\heffsize/2)} 
\pgfmathsetmacro{\ycoorb}{(\ycoor-\pixtocoord*\heffsize/2)} 
\pgfmathsetmacro{\result}{(\xcoorend+0.6*\scalh)} 
\ifx&#1&%
  \pgfmathsetmacro{\xcoorH}{\xcoorend}
\else
  \node at (\result,\ycoor){#1};
  \pgfmathsetmacro{\xcoorH}{(\xcoorend+0.25*\scalh)}
\fi
\ifnum#3=1
 \pgfmathsetmacro{\xxx}{((\xcoorend-\xcoorbeg)/2)} 
 \pgfmathsetmacro{\result}{\xcoorbeg}
\else
 \pgfmathsetmacro{\xxx}{((\xcoorend-\xcoorbeg)/( #3 - 1 ))} 
 \pgfmathsetmacro{\result}{(\xcoorbeg-\xxx)}
\fi
\foreach \x in {1,...,#3}
{
  \coordinate (vertex) at ($(\result,\ycoora)+(\x*\xxx,0)$);
  \node[inner sep=0pt,minimum size=0pt] (#4\x a) at  (vertex) {};
  \coordinate (vertex) at ($(\result,\ycoorb)+(\x*\xxx,0)$);
  \node[inner sep=0pt,minimum size=0pt] (#4\x b) at  (vertex) {};
  \coordinate (vtxvac) at ($(\result,\yvac)+(\x*\xxx,0)$);
  \node[inner sep=0pt,minimum size=0pt] (#4\x v1) at ($(vtxvac)-(\xvac,0)$) {};
  \node[inner sep=0pt,minimum size=0pt] (#4\x v2) at ($(vtxvac)+(\xvac,0)$) {};
  \coordinate (vtxvac) at ($(\result,\ycoord)+(\x*\xxx,0)$);
  \node[inner sep=0pt,minimum size=0pt] (#4\x vd1) at ($(vtxvac)-(\xvac,0)$) {};
  \node[inner sep=0pt,minimum size=0pt] (#4\x vd2) at ($(vtxvac)+(\xvac,0)$) {};
}
\node[inner sep=0pt,minimum size=0pt] (#4a) at  (#41a) {};
\node[inner sep=0pt,minimum size=0pt] (#4b) at  (#41b) {};
\node[inner sep=0pt,minimum size=0pt] (#4v1) at (#41v1) {};
\node[inner sep=0pt,minimum size=0pt] (#4v2) at (#41v2) {};
\node[inner sep=0pt,minimum size=0pt] (#4vd1) at (#41vd1) {};
\node[inner sep=0pt,minimum size=0pt] (#4vd2) at (#41vd2) {};
\ifx&#2&%
\else  
\pgfmathsetmacro{\result}{((\xcoorend+\xcoorbeg)/2)} 
\setcounter{stoer}{#2} 
\node[inner sep=0pt,minimum size=0pt] at (\result,\ycoor){{\footnotesize\slshape\sffamily \Roman{stoer}}}; 
\fi
}

\newcommand{\dCross}{$\mathbin{\tikz [x=1.4ex,y=1.4ex] \draw[thick] (0,0) -- (1,1) (0,1) -- (1,0);}$}

\newcommand{\dHone}[2][]{\pgfmathsetmacro{\xcoorbeg}{(\xcoorH+0.5*\scalh)}
\pgfmathsetmacro{\xcoorend}{(\xcoorbeg+0.5*\scalh)} 
\draw[hoper-line](\xcoorbeg,\ycoor) -- (\xcoorend,\ycoor);
\node at (\xcoorend,\ycoor){\dCross};
\pgfmathsetmacro{\result}{(\xcoorend+0.35*\scalh)} 
\ifx&#1&%
\pgfmathsetmacro{\xcoorH}{\xcoorend}
\else
\node at (\result,\ycoor){#1};
\pgfmathsetmacro{\xcoorH}{(\xcoorend+0.25*\scalh)}
\fi
\node[inner sep=0pt,minimum size=0pt] (#2) at (\xcoorbeg,\ycoor){}; 
\pgfmathsetmacro{\xx}{(\xcoorbeg-\xvac/2)} 
\node[inner sep=0pt,minimum size=0pt] (#2v1) at (\xx,\yvac) {}; 
\pgfmathsetmacro{\xx}{(\xcoorbeg+\xvac/2)} 
\node[inner sep=0pt,minimum size=0pt] (#2v2) at (\xx,\yvac) {};
\pgfmathsetmacro{\xx}{(\xcoorbeg-\xdvac/2)}
\node[inner sep=0pt,minimum size=0pt] (#2vd1) at (\xx,\ycoord) {};
\pgfmathsetmacro{\xx}{(\xcoorbeg+\xdvac/2)}
\node[inner sep=0pt,minimum size=0pt] (#2vd2) at (\xx,\ycoord) {};
\node[inner sep=0pt,minimum size=0pt] (#21) at  (#2) {};
\node[inner sep=0pt,minimum size=0pt] (#21v1) at (#2v1) {};
\node[inner sep=0pt,minimum size=0pt] (#21v2) at (#2v2) {};
}

\newcommand{\dHoner}[2][]{\pgfmathsetmacro{\xcoorbeg}{(\xcoorH+0.5*\scalh)}
\pgfmathsetmacro{\xcoorend}{(\xcoorbeg+0.5*\scalh)} 
\draw[hoper-line](\xcoorbeg,\ycoor) -- (\xcoorend,\ycoor);
\node at (\xcoorbeg,\ycoor){$\bf \times$};
\pgfmathsetmacro{\result}{(\xcoorend+0.35*\scalh)} 
\ifx&#1&%
\pgfmathsetmacro{\xcoorH}{\xcoorend}
\else
\node at (\result,\ycoor){#1};
\pgfmathsetmacro{\xcoorH}{(\xcoorend+0.25*\scalh)}
\fi
\node[inner sep=0pt,minimum size=0pt] (#2) at (\xcoorend,\ycoor){}; 
\pgfmathsetmacro{\xx}{(\xcoorend-\xvac/2)} 
\node[inner sep=0pt,minimum size=0pt] (#2v1) at (\xx,\yvac) {}; 
\pgfmathsetmacro{\xx}{(\xcoorend+\xvac/2)} 
\node[inner sep=0pt,minimum size=0pt] (#2v2) at (\xx,\yvac) {};
\pgfmathsetmacro{\xx}{(\xcoorend-\xdvac/2)}
\node[inner sep=0pt,minimum size=0pt] (#2vd1) at (\xx,\ycoord) {};
\pgfmathsetmacro{\xx}{(\xcoorend+\xdvac/2)}
\node[inner sep=0pt,minimum size=0pt] (#2vd2) at (\xx,\ycoord) {};
}

\newcommand{\dHtwo}[3][]{ 
\pgfmathsetmacro{\xcoorbeg}{(\xcoorH+0.5*\scalh)}
\pgfmathsetmacro{\xcoorend}{(\xcoorbeg+1*\scalh)} 
\draw[hoper-line](\xcoorbeg,\ycoor) -- (\xcoorend,\ycoor); 
\pgfmathsetmacro{\result}{(\xcoorend+0.35*\scalh)} 
\ifx&#1&%
\pgfmathsetmacro{\xcoorH}{\xcoorend}
\else
\node at (\result,\ycoor){#1};
\pgfmathsetmacro{\xcoorH}{(\xcoorend+0.25*\scalh)}
\fi
\node[inner sep=0pt,minimum size=0pt] (#2) at (\xcoorbeg,\ycoor) {};
\node[inner sep=0pt,minimum size=0pt] (#3) at (\xcoorend,\ycoor) {};
\pgfmathsetmacro{\xx}{(\xcoorbeg-\xvac/2)} 
\node[inner sep=0pt,minimum size=0pt] (#2v1) at (\xx,\yvac) {};
\pgfmathsetmacro{\xx}{(\xcoorbeg+\xvac/2)} 
\node[inner sep=0pt,minimum size=0pt] (#2v2) at (\xx,\yvac) {}; 
\pgfmathsetmacro{\xx}{(\xcoorend-\xvac/2)} 
\node[inner sep=0pt,minimum size=0pt] (#3v1) at (\xx,\yvac) {};
\pgfmathsetmacro{\xx}{(\xcoorend+\xvac/2)} 
\node[inner sep=0pt,minimum size=0pt] (#3v2) at (\xx,\yvac) {};
\pgfmathsetmacro{\xx}{(\xcoorbeg-\xdvac/2)} 
\node[inner sep=0pt,minimum size=0pt] (#2vd1) at (\xx,\ycoord) {};
\pgfmathsetmacro{\xx}{(\xcoorbeg+\xdvac/2)} 
\node[inner sep=0pt,minimum size=0pt] (#2vd2) at (\xx,\ycoord) {}; 
\pgfmathsetmacro{\xx}{(\xcoorend-\xdvac/2)} 
\node[inner sep=0pt,minimum size=0pt] (#3vd1) at (\xx,\ycoord) {};
\pgfmathsetmacro{\xx}{(\xcoorend+\xdvac/2)} 
\node[inner sep=0pt,minimum size=0pt] (#3vd2) at (\xx,\ycoord) {};
} 

\newcommand{\dHmany}[3][]{ 
\pgfmathsetmacro{\xcoorbeg}{(\xcoorH+0.5*\scalh)}
\pgfmathsetmacro{\xcoorend}{(\xcoorbeg+#2*\scalh/2)} 
\draw[hoper-line](\xcoorbeg,\ycoor) -- (\xcoorend,\ycoor); 
\pgfmathsetmacro{\result}{(\xcoorend+0.35*\scalh)} 
\ifx&#1&%
\pgfmathsetmacro{\xcoorH}{\xcoorend}
\else
\node at (\result,\ycoor){#1};
\pgfmathsetmacro{\xcoorH}{(\xcoorend+0.25*\scalh)}
\fi
\ifnum#2=1
 \pgfmathsetmacro{\xxx}{((\xcoorend-\xcoorbeg)/2)} 
 \pgfmathsetmacro{\result}{\xcoorbeg}
\else
 \pgfmathsetmacro{\xxx}{((\xcoorend-\xcoorbeg)/( #2 - 1 ))} 
 \pgfmathsetmacro{\result}{(\xcoorbeg-\xxx)}
\fi
\foreach \x in {1,...,#2}
{
  \coordinate (vertex) at ($(\result,\ycoor)+(\x*\xxx,0)$);
  \node[inner sep=0pt,minimum size=0pt] (#3\x) at  (vertex) {};
  \coordinate (vtxvac) at ($(\result,\yvac)+(\x*\xxx,0)$);
  \node[inner sep=0pt,minimum size=0pt] (#3\x v1) at ($(vtxvac)-(\xvac,0)$) {};
  \node[inner sep=0pt,minimum size=0pt] (#3\x v2) at ($(vtxvac)+(\xvac,0)$) {};
  \coordinate (vtxvac) at ($(\result,\ycoord)+(\x*\xxx,0)$);
  \node[inner sep=0pt,minimum size=0pt] (#3\x vd1) at ($(vtxvac)-(\xvac,0)$) {};
  \node[inner sep=0pt,minimum size=0pt] (#3\x vd2) at ($(vtxvac)+(\xvac,0)$) {};
}
\node[inner sep=0pt,minimum size=0pt] (#3) at  (#31) {};
\node[inner sep=0pt,minimum size=0pt] (#3v1) at (#31v1) {};
\node[inner sep=0pt,minimum size=0pt] (#3v2) at (#31v2) {};
\node[inner sep=0pt,minimum size=0pt] (#3vd1) at (#31vd1) {};
\node[inner sep=0pt,minimum size=0pt] (#3vd2) at (#31vd2) {};
} 

\newcommand{\dTs}[3][]{\dAmp{#1}{#2}{#3}}

\newcommand{\dTds}[3][]{\dAmpD{#1}{#2}{#3}}

\newcommand{\dTone}[3][]{\dAmp[$_{T_1}$]{#1}{1}{#2}; \draw (#2) node[below] {\footnotesize\slshape\sffamily #3};}
\newcommand{\dTtwo}[4][]{\dAmp[$_{T_2}$]{#1}{2}{#2}; \draw (#21) node[below] {\footnotesize\slshape\sffamily #3};\draw (#22) node[below] {\footnotesize\slshape\sffamily #4};}
\newcommand{\dTsone}[3][]{\dAmp{#1}{1}{#2}; \draw (#2) node[below] {\footnotesize\slshape\sffamily #3};}
\newcommand{\dTstwo}[4][]{\dAmp{#1}{2}{#2}; \draw (#21) node[below] {\footnotesize\slshape\sffamily #3};\draw (#22) node[below] {\footnotesize\slshape\sffamily #4};}

\newcommand{\dTdsone}[3][]{\dAmpD{#1}{1}{#2}; \draw (#2) node[above] {\footnotesize\slshape\sffamily #3};}
\newcommand{\dTdstwo}[4][]{\dAmpD{#1}{2}{#2}; \draw (#21) node[above] {\footnotesize\slshape\sffamily #3};\draw (#22) node[above] {\footnotesize\slshape\sffamily #4};}
\newcommand{\dTdvone}[3][]{
  \tikzset{exoper-line/.style={exvac-line-save}}
  \ifx&#1&%
    \dTdsone{#2}{#3}
  \else
    \dTdsone[#1]{#2}{#3}
  \fi
  \tikzset{exoper-line/.style={exoper-line-save}} }
\newcommand{\dTdvtwo}[4][]{
  \tikzset{exoper-line/.style={exvac-line-save}}
  \ifx&#1&%
    \dTdstwo{#2}{#3}{#4}
  \else
    \dTdstwo[#1]{#2}{#3}{#4}
  \fi
  \tikzset{exoper-line/.style={exoper-line-save}} }
\newcommand{\dUone}[3][]{
  \tikzset{exoper-line/.style={exoper2-line-save}}
  \ifx&#1&%
    \dTone{#2}{#3}
  \else
    \dTone[#1]{#2}{#3}
  \fi
  \tikzset{exoper-line/.style={exoper-line-save}} 
}
\newcommand{\dUtwo}[4][]{
  \tikzset{exoper-line/.style={exoper2-line-save}}
  \ifx&#1&%
    \dTtwo{#2}{#3}{#4}
  \else
    \dTtwo[#1]{#2}{#3}{#4}
  \fi
  \tikzset{exoper-line/.style={exoper-line-save}} 
}
\newcommand{\dUsone}[3][]{
  \tikzset{exoper-line/.style={exoper2-line-save}}
  \ifx&#1&%
    \dTsone{#2}{#3}
  \else
    \dTsone[#1]{#2}{#3}
  \fi
  \tikzset{exoper-line/.style={exoper-line-save}} 
}
\newcommand{\dUstwo}[4][]{
  \tikzset{exoper-line/.style={exoper2-line-save}}
  \ifx&#1&%
    \dTstwo{#2}{#3}{#4}
  \else
    \dTstwo[#1]{#2}{#3}{#4}
  \fi
  \tikzset{exoper-line/.style={exoper-line-save}} 
}

\newcommand{\dFs}[1]{\dHone{#1}}

\newcommand{\dWs}[2]{\dHtwo{#1}{#2}}

\newcommand{\dline}[3][]{
\ifx&#1&%
\draw[ph-line](#2) -- (#3);
\else
\draw[ph-line](#2) -- node[inner sep=0pt,minimum size=0pt,right = 0.5pt] {\footnotesize\slshape\sffamily #1} (#3);
\fi
}
\newcommand{\dcurve}[3][]{
\ifx&#1&%
\draw[ph-line](#2) to [bend right=30] (#3);
\else
\draw[ph-line](#2) to [bend right=30] node[inner sep=0pt,minimum size=0pt,right = 0.5pt] {\footnotesize\slshape\sffamily #1} (#3);
\fi
}
\newcommand{\dcurver}[3][]{
  \tikzset{ph-line/.style={ph-liner-save}}
  \dcurve[#1]{#3}{#2}
  \tikzset{ph-line/.style={ph-line-save}}
}

\newcommand{\dmoveT}[1]{\pgfmathsetmacro{\xcoor}{(\xcoor+#1*0.25*\scalh)}}
\newcommand{\dmoveTd}[1]{\pgfmathsetmacro{\xcoordg}{(\xcoordg+#1*0.25*\scalh)}}

\newcommand{\dsavediags}[2][]{
 \ifx&#1&%
   \newcommand{\DiagramName}{Diag}
 \else
   \newcommand{\DiagramName}{#1}
 \fi 
 \newcounter{DiagCounter}
 \pgfrealjobname{#2}
}

\newcommand{\diagsav}[2][]{
 \ifx&#1&%
  \stepcounter{DiagCounter}
  \beginpgfgraphicnamed{\DiagramName\theDiagCounter}{#2}
  \endpgfgraphicnamed
 \else
  \beginpgfgraphicnamed{#1}{#2}
  \endpgfgraphicnamed
 \fi
} 

\definecolor{orcidlogocol}{HTML}{A6CE39}
\tikzset{
  orcidlogo/.pic={
    \fill[orcidlogocol] svg{M256,128c0,70.7-57.3,128-128,128C57.3,256,0,198.7,0,128C0,57.3,57.3,0,128,0C198.7,0,256,57.3,256,128z};
    \fill[white] svg{M86.3,186.2H70.9V79.1h15.4v48.4V186.2z}
                 svg{M108.9,79.1h41.6c39.6,0,57,28.3,57,53.6c0,27.5-21.5,53.6-56.8,53.6h-41.8V79.1z M124.3,172.4h24.5c34.9,0,42.9-26.5,42.9-39.7c0-21.5-13.7-39.7-43.7-39.7h-23.7V172.4z}
                 svg{M88.7,56.8c0,5.5-4.5,10.1-10.1,10.1c-5.6,0-10.1-4.6-10.1-10.1c0-5.6,4.5-10.1,10.1-10.1C84.2,46.7,88.7,51.3,88.7,56.8z};
  }
}

\newcommand\orcidicon[1]{\href{https://orcid.org/#1}{\mbox{\scalerel*{
\begin{tikzpicture}[yscale=-1,transform shape]
\pic{orcidlogo};
\end{tikzpicture}
}{|}}}}

\begin{document}


\title[]{
Highly Accurate Expectation Values Using High-Order Relativistic Coupled Cluster Theory}
\author{Gabriele Fabbro \orcidicon{0009-0009-7625-7906}}
\author{Jan Brandejs \orcidicon{0000-0002-2107-3095}}
\author{Trond Saue \orcidicon{0000-0001-6407-0305}}\email{trond.saue@irsamc.ups-tlse.fr}
\homepage{https://dirac.ups-tlse.fr/saue}
\affiliation{Laboratoire de Chimie et Physique Quantique,\\UMR 5626 CNRS - Université Toulouse III-Paul Sabatier,\\ 118 Route de Narbonne, F-31062 Toulouse, France}

\date{\today}
             
\begin{abstract}
\section*{Abstract}
This work presents the automatic generation of analytic first derivatives of the energy for general coupled-cluster models using the \text{tenpi} toolchain.
We report the first implementation of expectation values for CCSDT and CCSDTQ methods within the DIRAC program package for relativistic molecular calculations.
As pivotal calculations, we focus on the electric field gradient (EFG) evaluated at the lithium nucleus in LiX (X = H, F, Cl) compounds, enabling the extraction of the nuclear electric quadrupole moment $Q(\ce{^{7}Li})$, and at the aluminum nucleus in AlY (Y =H, F, Cl, Br) compounds, for the determination of $Q(\ce{^{27}Al})$. These high-order methods are applied to compute corrections for triple and quadruple excitations for the EFG, a crucial quantity for determining nuclear quadrupole moments. We obtain \( Q(\ce{^{27}Al}) = 0.1466 \)\,b, in excellent agreement with the recommended value, and \( Q(\ce{^7Li}) = -0.0386 \)\,b, which is smaller than the currently recommended value, that indicates the need for further investigation.

\end{abstract}

\maketitle

\section{\label{sec:level1}Introduction}


The atomic nucleus, while often simplified as a point charge in basic atomic models, possesses a complex internal structure that can manifest in various electromagnetic moments. Among these, the nuclear electric quadrupole moment (NQM) stands out as a fundamental property that describes the deviation of the nuclear charge distribution from perfect spherical symmetry. This characteristic is inherent to nuclei with a spin quantum number greater than one-half ($I > 1/2$). The determination of the nuclear quadrupole moment plays a crucial role in advancing our understanding of nuclear structure and the distribution of charge within nuclei.\cite{Gordy_PR_1949,Townes_PR_1949,Moszkowski_PR_1954,Marshalek_RMP_1963,SHARON1967321,Neugart_Neyens_2006,Stone_Int2024} One of the key mechanisms for the accurate determination of the (spectroscopic-)nuclear quadrupole moment is based on the interaction between the NQM and the electric field gradient (EFG) at the nuclear position, and it can be quantified by the nuclear quadrupole coupling constant (NQCC), defined as 
\begin{equation}
\label{NQCC}
    \text{NQCC [in MHz]}=234.9647\,\times\,Q\text{[in b]}\times q\,\left[\text{in }E_{h}/a_{0}^{2}\right]
\end{equation}
where \( eQ \) represents the nuclear electric quadrupole moment of the isotope and \(eq \) denotes the electric field gradient along the molecular axis at the location of the nucleus, computed using molecular electronic structure theory. Several experimental techniques exist for the accurate determination of NQCCs, which have been cataloged by Stone,\cite{Stone_Int2024} and on which we recently performed a statistical analysis.\cite{Fabbro_JPCA_2025} On the other hand, the accurate determination of the EFG is challenging, since it is quite sensitive to the basis set quality, as clearly show by van Stralen and Vissher.\cite{VANSTRALEN10072003} Moreover, at the SCF-level, in the absence of polarization, the EFG is zero for core shells. Instead, it samples the inner tails of valence orbitals, making it sensitive to electron correlation effects.

Coupled Cluster (CC) theory \cite{CrawfordTDaniel2000AItC,BartlettRev20007} is highly regarded as the "gold standard" in quantum chemistry\cite{Bartlett1984,Helgaker2004} because it can systematically capture electronic correlation.\cite{Lowdin} 
Moreover, the inherent multiplicative-separable characteristic of the CC wave function leads to size-extensivity and size-consistency,\cite{Paldus1972,Pople1978,Pal1984,HANRATH200931,CrawfordTDaniel2000AItC,BartlettRev20007} an essential feature for accurately describing realistic systems. 
The computation of properties within the CC framework can be traced back to 1969 by Čížek,\cite{Cížek1969} where he showed how expectation values can be computed by factorizing the numerator in order to remove all the disconnected diagram contributions. However, the expression provided by Čížek was problematic due to the enormous quantity of terms, since there was not a natural truncation of the cluster expansion. A few years later Monkhorst\cite{Monkhorst1977} provided a strategy for the computation of properties (both time dependent and independent) with the usage of the similarity transformed Hamiltonian, which provides, using the Baker–Campbell–Hausdorff expansion, a natural truncation of terms, still completely connected. 
In the 1980s, analytic energy gradients were developed by Adamowicz and Bartlett.\cite{Adamowicz1984} Bartlett and co-workers provided general diagrammatic and algebraic expressions for the analytical derivatives of CC and MBPT wave functions. \cite{Fitzgerald1986_JPC} After a few years, Salter and co-workers\cite{Salter1987}  showed the expressions for the energy derivatives and response density for the CCSD method.\cite{Purvis1982}  Until then, the bottleneck in the calculation of molecular properties at the CC level was the requirement for amplitude derivatives, and thus the need to solve the response equations for each field strength. 
Salter and co-workers showed that it is possible to rearrange the CC equations to yield a single set of linear, perturbation-independent equations, namely \textit{lambda equations}.\cite{Salter1989}  Helgaker and Jørgensen provided a more natural framework to recover these equations.\cite{Jorgensen1988,Helgaker1989} They observed that by introducing a Lagrangian and by enforcing its stationarity with respect to the field strengths, it becomes possible to get the lambda equations, and therefore calculate first- and second-order molecular properties without the need to explicitly compute the derivatives of the cluster amplitudes (\textit{vide infra}). The Lagrangian approach has been a fertile ground for the calculation of molecular properties at the CC level, both for ground\cite{Koch1990,Gauss1991,Gauss1995}  and excited states.\cite{Koch1990_a,Koch1990_b,Stanton1993,Stanton:1993vcu}
Later on, Kállay and co-workers developed analytic first and second derivatives for general coupled-cluster excitation level, based on the Lagrangian formalism.\cite{Kallay2003,Kallay2004} 

In addition to correlation effects, the EFG is highly sensitive to relativistic effects,\cite{KELLO1990641,Pyykkoe_TCA1997,Visscher1998,Pernpointner2004,Pyykkö1988,Saue2011,Pyykko2012,Pyykkö2012} as it is proportional to the inverse cubic radial expectation value, $\braket{r^{-3}}$ (see, for instance, Ref.\citenum{TSAUEBOOK}). This proportionality highlights that the EFG probes the core region surrounding the nucleus, where relativistic effects are particularly pronounced, especially for heavy elements (typically with \( Z > 40 \)). The increased number of electrons in heavy atoms further amplifies the importance of electronic correlation effects, making both relativistic and correlation contributions critical for accurately determining the EFG. The primary challenge in studying systems with heavy atoms is that electron correlation effects and relativistic effects are non-additive and must be treated simultaneously.\cite{Autschbach2012}  

Although alternative approaches incorporate relativistic effects via perturbation theory,\cite{Cheng2011,Cheng2011_2,Cheng2012} the most natural inclusion is obtained by using a four‐component Dirac–Coulomb (DC) Hamiltonian.  In the \text{DIRAC} program package\cite{DIRAC2020} first-order properties at the relativistic CCSD level were implemented by Shee and co-workers\cite{Shee2006} in the \text{RELCCSD} module,\cite{Visscher1995,Visscher1996,Visscher2001} which benefits from point-group symmetry.\cite{DIRAC2020} More recently, Pototschnig and co-workers implemented relativistic coupled-cluster algorithms, including the CCSD analytic gradient, optimized for modern heterogeneous high-performance computing infrastructures, with support for GPU co-processing via the \text{ExaTENSOR} library.\cite{Pototschnig2021} 

Contributions from excitations beyond doubles may be required for an accurate description of the target property. An example of that is provide by the determination of the NQM of $\ce{^{27}Al}$, where Pernpointner and Visscher showed that the inclusion fo the triple excitations can still effect the EFG.\cite{Pernpointner2001_JCP} Stopkowicz and Gauss also showed how triple and quadruple excitations can affect the NQM of sulfur isotopes.\cite{PhysRevA.90.022507} 
Therefore, the development of a general-order program arose for the generation of high-order CC equations. This need stems from the complexity and difficulty of explicitly programming and debugging the various matrix elements associated with excitations higher than doubles. Several approaches can be employed to derive  systematically the CC equations: by directly applying second-quantization rules,\cite{D4CP00444B,liebenthal2025} some rely on Wick's theorem,\cite{Wick_PR_1950} which decomposes products of operators into sums of contractions,\cite{KENDALL2000260,Kohn2008,Neuscamman2009,Saitow2013,Song2022,Bochevarov2004,Quintero-Monsebaiz_AIP_2023} while others utilize a diagrammatic approach,\cite{Shavitt_Bartlett_2009} where the equations are derived graphically.\cite{Kállay2001,Shiozaki2018} The major advantage of using diagrams over second-quantized algebra and Wick's theorem is that they are easier to inspect and interpret, and are therefore less prone to errors. Moreover, it is more easy to generate unique diagrams and identify those that are  equivalent. Recently, Brandejs and co-workers introduced \text{tenpi},\cite{brandejs2024generatingcoupledclustercode,tenpiGitlab} an open-souce toolchain designed for the development of CC methods, integrated within the DIRAC program package.\cite{DIRAC2020} \text{tenpi} builds upon the diagrammatic approach of Kállay and Surján,\cite{Kállay2001} which is in turn based on the Kucharski-Bartlett diagrams,\cite{Kucharski1992} but extends it with advanced features such as global optimization of intermediates, a Python-based user interface, a visualization module, and a code generator. On the output, \text{tenpi} supports several tensor libraries including ExaTENSOR as well as the standard tensor interface TAPP (Tensor Algebra Processing Primitives).\cite{tappStandardInterfaceGithub}

In this work, we developed and implemented high-order relativistic CC expectation values within the DIRAC program package,\cite{DIRAC2020} focusing on the CCSDT and CCSDTQ methods. The equations were derived and generated using the \text{tenpi} toolchain and subsequently implemented in the ExaCorr module in DIRAC.\cite{Pototschnig2021} To validate our protocol, we performed non-relativistic calculations and compared the results with those obtained using the well-established MRCC program.\cite{kallay2020mrcc} Furthermore, as sample applications, we computed the EFG in the LiX (X=H, F, Cl) and AlY (Y=H, F, Cl, Br) series of systems to extract the NQMs of \ce{^{7}Li} and \ce{^{27}Al}, comparing these with the standard reference values.\cite{Pyykko_MP2001,Pyykkö2008_MP}

The paper is organized as follows: In Sec.\ref{sec:level2} we will outline the theory behind the computation of time-independent molecular properties using a relativistic-coupled-cluster wave-function. We will then discuss the generation of the lambda and density matrix equations in Sec.\ref{sec:level3}, arriving at Sec.\ref{sec:level4} where the computational protocol is described. Subsequently, in Sec.\ref{sec:level5} sample applications concerning the determination of the NQMs of \ce{^{7}Li} and \ce{^{27}Al} will be discussed, followed by Sec.\ref{sec:level6} where final conclusions and future perspectives are given.

Throughout this work, we use SI units unless otherwise stated.

\section{\label{sec:level2}Theory}

\subsection{\label{sec:level2.1}Relativistic framework}
Relativistic molecular calculations are performed using the electronic Hamiltonian within the Born-Oppenheimer approximation
\begin{equation}
\label{eq:hamiltonian}
\hat{H}=\sum_{i}\hat{h}_{{D}}(i)+\sum_{i<j}\hat{g}(i,j)+V_{\text{NN}}; \quad V_{\text{NN}}=\frac{e^{2}}{4\pi\varepsilon_{0}}\sum_{A<B}\frac{Z_{A}Z_{B}}{R_{AB}},
\end{equation}
where the mono-electronic part, $\hat{h}_{{D}}$, is the Dirac Hamiltonian
\begin{equation}
    \hat{h}_{{D}}(i)=c\left(\bm{\alpha}_{i}\cdot\mathbf{p}_{i}\right)+{\beta}_{i}m_{e}c^{2}+V_{eN}(r_{i}),
\end{equation}
and $\bm{\alpha}=(\alpha_{x},\alpha_{y},\alpha_{z})$ and ${\beta}$ are the the Dirac matrices, $\mathbf{p}$ is the linear momentum, $m_{e}$ is the mass of the electron and $c$ is the speed of light ($c=$ 137.0359998 $a_{0} E_{h}/\hbar$). The electron-nucleus interaction $V_{eN}$ is defined as 
\begin{equation}
    V_{eN}(r)=-e\sum_{A}\varphi_{A}(r), \quad \varphi_{A}(r_{1})=\frac{1}{4\pi\varepsilon_{0}}\int\frac{\rho_{A}(r_{2})}{|\mathbf{r}_{1}-\mathbf{r}_{2}|}d^{3}\mathbf{r}_{2},
\end{equation}
where $e$ is the fundamental charge, $\varepsilon_{0}$ is the electric constant, $\varphi_{A}$ and $\rho_{A}$ are respectively the scalar potential and  the charge density of nucleus $A$. In our case, the bi-electronic term $\hat{g}(i,j)$ is given by the instantaneous Coulomb interaction $\hat{g}^{\text{C}}(i,j)$ and the Gaunt interaction $\hat{g}^{\text{G}}(i,j)$ 
\begin{equation}
    \hat{g}(i,j) = \hat{g}^{\text{C}}(i,j)+\hat{g}^{\text{G}}(i,j)=\frac{e^{2}}{4\pi\epsilon_0} \left( \frac{1}{r_{ij}} - \frac{c\alpha_i \cdot c\alpha_j}{c^2r_{ij}} \right).
\end{equation}
which describes the charge-charge and the current-current instantaneous interactions between the electrons in the chosen reference frame.\cite{saue2003four} Therefore, we will refer to the Hamiltonian in Eq.~\eqref{eq:hamiltonian} as the Dirac-Coulomb-Gaunt Hamiltonian (DCG). The eigenstates of the Dirac-Hamiltonian are four-component wave functions
\begin{equation}
    \hat{h}_{D}\begin{bmatrix}
        \psi^{\text{L}}\\
        \psi^{\text{S}}
    \end{bmatrix}=E\begin{bmatrix}
        \psi^{\text{L}}\\
        \psi^{\text{S}}
    \end{bmatrix}, \quad \psi^{L}=\begin{bmatrix}
        \psi^{\text{L}\alpha}\\
        \psi^{\text{L}\beta}
    \end{bmatrix}, \quad \psi^{\text{S}}=\begin{bmatrix}
        \psi^{\text{S}\alpha}\\
        \psi^{\text{S}\beta}
    \end{bmatrix}
\end{equation}
where we have defined the large and the small components, $\psi^{\text{L}}$ and $\psi^{\text{S}}$, respectively. The use of a 4-component relativistic theory presents significant challenges, such as the presence of negative energy orbitals which imply that the electronic Hamiltonian does not have bound-state solutions, as pointed out by Brown and Ravenhall.\cite{Brown1951} However, by employing the no-pair approximation,\cite{Sucher1980} the molecular orbitals are optimized at the Hartree-Fock (HF) level, and only the positive-energy ones are retained at the correlated level. The large and small components of the spinor are expanded in basis sets, $\{{\chi}^{\text{L}}\}$ and $\{{\chi}^{\text{S}}\}$, that satisfy the kinetic-balance prescription.\cite{Stanton1984} 

 While the four-component formulation is highly accurate, it is computationally demanding. Two-component methods offer a significant cost reduction at the SCF level by retaining key relativistic effects, but this advantage does not necessarily extend to correlated calculations. Various approaches have been developed to decouple the solutions with positive energy from those with negative energy. Some of them are based on the so-called 'elimination technique',\cite{Dyall1997,Filatov2005,Filatov2006,Filatov2007} while others rely on unitary decoupling transformations,\cite{Foldy_PR_1950,Barysz2001,Douglas1974,Hess1986,Hess1985,Chang1986,Lenthe1993} but the two approaches are entirely equivalent.\cite{Iliaš2007_JCP} 
Prominent among these is the eXact 2-component (X2C) Hamiltonian, which is an exact decoupling of the positive and negative energy solutions.\cite{jensen:rehe2005,kutzelnigg:jcp2005,Iliaš2007,liu:jcp2009} While the one-electron part of the Hamiltonian admits a relatively straightforward decoupling through unitary transformations or algebraic techniques, the two-electron component presents significant computational challenges. 
The X2C transformation would require the evaluation of the full set of four-component two-electron integrals and the application of the decoupling transformation, making it potentially more demanding than the original four-component calculation. At the correlated level, the molecular-mean-field (X2Cmmf) Hamiltonian can be adopted.\cite{Sikkema2009} For correlated calculations, it is often advantageous to work within the framework of the normal-ordered Hamiltonian,
\begin{equation}
\label{normal-ordered-hamiltonian}
\hat{H}_{N} = \sum_{pq} f_{pq} \{a_p^\dagger a_q\} + \frac{1}{4} \sum_{pqrs}g_{pqrs}\{a_p^\dagger a_q^\dagger a_s a_r\},
\end{equation}
 where\(f_{pq}\) are the matrix elements of the Fock operator, ${g}_{pqrs}$ are the antisymmetrized two-electron integrals and the curly brackets indicate normal ordering with respect to the chosen reference state. Thus, in the X2cmmf approach, a four-component SCF calculation is first performed, and the decoupling transformation is applied to the Fock operator, hence eliminating picture-change errors in the mean-field description at the correlated level. The two-component Fock matrix is combined with the untransformed two-electron interaction, and the correlated calculation is carried out.
 For more details, see for instance Ref.\citenum{Saue2011}.

\subsection{\label{sec:level2.2}Coupled Cluster Theory}
\label{CC_chapter}
Coupled Cluster theory has been well-established for many years, and we suggest the comprehensive reviews made by Crawford and Schaefer\cite{CrawfordTDaniel2000AItC},  Bartlett and Shavitt\cite{BartlettRev20007} as well the CC chapter in the ESQC book\cite{crawford2025cc} for further insights and details. 

In Coupled Cluster (CC) theory, the correlated wave function is defined through a nonunitary exponential parameterization with respect to a certain reference $\ket{\Phi_{0}}$, which often is the Hartree-Fock determinant
\begin{equation}
    \ket{\Psi_{\text{CC}}}=e^{\hat{T}}\ket{\Phi_{0}}, \quad \hat{T}=\sum_{\mu}t_{\mu}\hat{\tau}_{\mu},
\end{equation}
and where $\hat{T}$ is the cluster operator, expressed as a combination of excitation operators $\hat{\tau}_{\mu}$.\cite{Bartlett1989} In practice,  the excitation operators are restricted to act only to some maximal excitation level $k_{\text{max}}$, giving rise the commonly known CC approximations.\cite{Purvis1982,Lee1984,Urban1985,Noga1987,Kucharski1992,Musiał2002} Therefore, the truncated cluster operator reads as
\begin{equation}
    \hat{T}=\sum_{k}^{k_{\text{max}}}\hat{T}_{k}, \quad \hat{T}_{k}=\sum_{\mu}t_{\mu;k}\hat{\tau}_{\mu;k}.
\end{equation}
The amplitudes ${{t}_{\mu;k}}$ are found by inserting the CC ansatz into the wave-equation
\begin{equation}
\label{wave-eq}
    \hat{H}_{N}\ket{\Psi_{\text{CC}}}=E_{\text{CC}}\ket{\Psi_{\text{CC}}}.
\end{equation}
After a pre-multiplication on the left by $e^{-\hat{T}}$ and by left-projecting Eq.~\eqref{wave-eq} with the reference $\bra{\Phi_{0}}$ and the excited determinants $\bra{\Phi_{\mu;k}}$, the coupled cluster equations are provided
\begin{align}
\label{eqn_energy}
    E_{\text{CC}}          &= \braket{\Phi_{0}|\hat{\overline{H}}|\Phi_{0}}, 
    & \quad \hat{\overline{H}} &= e^{-\hat{T}} \hat{H}_{N} e^{\hat{T}}, \\
\label{eqn_amplitudes}
    0                      &= \braket{\Phi_{\mu;k}|\hat{\overline{H}}|\Phi_{0}}, 
    & \quad \ket{\Phi_{\mu;k}} &= \hat{\tau}_{\mu;k} \ket{\Phi_{0}}, \quad k \leq k_{\text{max}}
\end{align}

Eq.~\eqref{eqn_energy} is the equation for the energy, whereas Eq.~\eqref{eqn_amplitudes} are the equations for the amplitudes. 
This projective formulation employs the similarity-transformed Hamiltonian, $\hat{\overline{H}}$, which can be expanded using the Baker–Campbell–Hausdorff formula as
\begin{align}
\begin{split}
\label{bch}
    \hat{\overline{H}} &= \hat{H}_{N} + [\hat{H}_{N}, \hat{T}] + \frac{1}{2!} [[\hat{H}_{N}, \hat{T}], \hat{T}] \\
    &\quad + \frac{1}{3!} [[[ \hat{H}_{N}, \hat{T}], \hat{T}], \hat{T}] 
    + \frac{1}{4!} [[[[\hat{H}_{N}, \hat{T}], \hat{T}], \hat{T}], \hat{T}].
\end{split}
\end{align}
 which leads to only five non-zero terms, showing the advantage of the similarity-transformed Hamiltonian. As pointed out by Koch and Jørgensen,\cite{Koch1990_a} summing up Eq.~\eqref{eqn_energy} and Eq.~\eqref{eqn_amplitudes} we get
\begin{align}
\label{left}
    &\braket{\Phi_{\text{L}}|\hat{\overline{H}}|\Phi_{\text{R}}}=E_{\text{CC}}, \quad \ket{\Phi_{\text{R}}}=\ket{\Phi_{0}}, \quad \bra{\Phi_{\text{L}}}=\bra{\Phi_{0}}+\sum_{\mu,k}\overline{t}_{\mu;k}\bra{\Phi_{\mu;k}},\\
    &\braket{\Phi_{\text{L}}|\Phi_{\text{R}}}=1,
\end{align}
which clearly shows the non-hermitian nature of the similarity-transformed Hamiltonian, since it has different right- and left- eigenvectors. The coefficients $\overline{t}_{\mu;k}$ can be in principle determined by solving the left-eigenvalue problem
\begin{equation}
\label{left-eigen}
    \bra{\Phi_{L}}\hat{\overline{H}}=E_{\text{CC}}\bra{\Phi_{L}},
\end{equation}
by projecting it onto $\ket{\Phi_{\mu;k}}$
\begin{equation}
\label{projected-left}
    \braket{\Phi_{L}|\hat{\overline{H}}|\Phi_{\mu;k}}=E_{\text{CC}}\overline{t}_{\mu;k}
\end{equation}
where we also see that we get an energy-dependent equation. However, since left and right eigenvectors share the same eigenvalue (the CC energy) we may simply consider the right-eigenvalue problem. As we will see in the next section, the left-eigenvalue problem is connected to the computation of properties. 

It can be shown that the only terms contributing to both Eq.~\eqref{eqn_energy} and (\ref{eqn_amplitudes}) are those in which the Hamiltonian has at least one contraction with each cluster operator on its right. In the diagrammatic representation, this corresponds to diagrams where the Hamiltonian has at least one connection to each cluster operator.

Currently, Eq.~\eqref{eqn_energy} and Eq.~\eqref{eqn_amplitudes} can be expanded from \text{tenpi},\cite{brandejs2024generatingcoupledclustercode} up to arbitrary order. In the application part, we will restrict ourself to at maximum quadruple excitations
\begin{align}
    \hat{T}&=\hat{T}_{1}+\hat{T}_{2}+\hat{T}_{3}+\hat{T}_{4}\\
    \hat{T}_{1}&=\sum_{ia}t_{i}^{a}a_{a}^{\dagger}a_{i}\\
    \hat{T}_{2}&=\left(\frac{1}{2}\right)^{2}\sum_{ijab}t_{ij}^{ab}a_{a}^{\dagger}a_{b}^{\dagger}a_{j}a_{i}\\
     \hat{T}_{3}&=\left(\frac{1}{3!}\right)^{2}\sum_{ijkabc}t_{ijk}^{abc}a_{a}^{\dagger}a_{b}^{\dagger}a_{c}^{\dagger}a_{k}a_{j}a_{i}\\
     \hat{T}_{4}&=\left(\frac{1}{4!}\right)^{2}\sum_{ijklabcd}t_{ijkl}^{abcd}a_{a}^{\dagger}a_{b}^{\dagger}a_{c}^{\dagger}a_{d}^{\dagger}a_{l}a_{k}a_{j}a_{i}.
\end{align}

\subsection{\label{sec:level2.3} Time-independent Molecular Properties using Coupled Cluster theory}
\label{tdmpcc}
For a deeper understanding of the theory related to the calculation of molecular properties, we suggest the books by Saue,  Norman and Ruud,\cite{TSAUEBOOK} as well as  Sauer.\cite{Sauer2011} Additionally, the review by Helgaker and co-workers is recommended for the general theory and the historical development of molecular properties in electronic structure theory.\cite{Helgaker2012_CR}

We will derive the time-independent molecular properties within the framework of variational perturbation theory.\cite{Helgaker1992} The starting point is the introduction of a time-independent perturbation $\hat{V}(\bm{\varepsilon})$ that generate the perturbed Hamiltonian $\hat{H}(\bm{\varepsilon})$ from the unperturbed system, defined by \(\hat{H}_{0}\)
\begin{equation}
\label{pert_H}
    \hat{H}(\bm{\varepsilon})=\hat{H}_{0}+\hat{V}(\bm{\varepsilon}), \quad \hat{V}(\bm{\varepsilon})=\sum_{X}\varepsilon_{X}\hat{H}_{X}
\end{equation}
where $\bm{\varepsilon}$ is a collection of field-strengths. The unperturbed Hamiltonian $\hat{H}_{0}$ in the present work is the Dirac-Coulomb-Gaunt Hamiltonian defined in Eq.~\eqref{eq:hamiltonian}. 
 Variational perturbation theory is based on an optimization of the energy at each field strength
\begin{equation}
\label{varpertEq}
    \frac{dE(\bm{\varepsilon},\bm{\lambda})}{d\bm{\lambda}}\Bigg|_{\bm{\varepsilon}}=0.
\end{equation}
This condition implies that the variational parameters $\bm{\lambda}$ depend on the perturbation strength, expressed as $\bm{\lambda}\equiv\bm{\lambda}(\bm{\varepsilon})$.  Thus, first-order energy derivatives are computed by invoking the chain rule

\begin{equation}
    E^{[1]}=\frac{dE(\bm{\varepsilon},\bm{\lambda})}{d{\varepsilon}_{\text{A}}}=\left[{\frac{\partial E}{\partial {\varepsilon}_{\text{A}}}}+\sum_{i}{\frac{\partial E}{\partial{\lambda}_{i}}\frac{d{\lambda}_{i}}{d{\varepsilon}_{\text{A}}}}\right].
\end{equation}
We obtained the explicit derivative of the energy with respect to the field strength and an implicit derivative that includes the variation of the wave function parameters with respect to the field strength, often referred to as the response term. Typically, computing the response term is quite demanding because it requires solving a set of response equations, and the number of these equations depends on the number of field strengths. However, for fully-variational wave functions, first-order energy derivatives can be easily computed using the Hellmann–Feynman theorem\cite{Hellmann1937,Feynman1938,Olsen1985,Christiansen1998} since the energy is variationally optimized with respect to the variational parameters, making the implicit term zero. As noted independently by Hylleraas\cite{Hylleraas1930} and Wigner,\cite{Wigner1935} using Eq.~\eqref{varpertEq}, we can derive the $(2n+1)$ rule: knowing the $n$-th derivative of the parameters allows us to determine the $(2n+1)$-th order derivative of the energy.

The situation is less trivial for non-variational wave functions, since the energy is not variationally optimized respect to the parameters, which suggest that we should compute the response term for each field strength. To overcome this obstacle, Helgaker and Jørgensen\cite{Jorgensen1988,Helgaker1989,Koch1990} considered a constrained optimization process of a CC Lagrangian function, where the constrains are Eqs.(\ref{eqn_amplitudes})
\begin{align}
\begin{split}
\label{lagrangian}
    L_{\text{CC}}(\bm{\varepsilon},\bm{t},\overline{\bm{t}})&=E_{\text{CC}}+\sum_{\eta,k}\overline{t}_{\eta;k}\braket{\Phi_{\eta;k}|\hat{\overline{H}}|\Phi_{0}}, \quad k\le k_{\text{max}}
    \end{split}
\end{align}
By simply defining a de-excitator operator $\hat{\Lambda}$ as
\begin{equation}
    \hat{\Lambda}=\sum_{k}^{k_{\text{max}}}\hat{\Lambda}_{k}, \quad \hat{\Lambda}_{k}=\sum_{\eta}\overline{t}_{\eta;k}\hat{\tau}_{\eta;k}^{\dagger},
\end{equation}
defined as a linear combination of de-excitator operators $\hat{\tau}_{\eta;k}^{\dagger}$ associated to the de-excitation manifold corresponding to the cluster operator $\hat{T}$, and $\overline{{t}}_{\eta;k}$ are the multipliers, we see that Eq.~\eqref{lagrangian} may be written as
\begin{equation}
     L_{\text{CC}}(\bm{\varepsilon},\bm{t},\overline{\bm{t}})=\braket{\Phi_{0}|(1+\hat{\Lambda})\hat{\overline{H}}|\Phi_{0}}=\braket{\Phi_{\text{L}}|\hat{\overline{H}}|\Phi_{\text{R}}},
\end{equation}
which is exactly Eq.~\eqref{left}.

We should note that this definition of the Lagrangian overlooks orbital relaxation, but which is assumed to be partially accounted for by the $\hat{T}_{1}$ operator.\cite{Thouless1960,Salter1987} Imposing the stationarity of the Lagrangian with respect to both the multipliers and the amplitudes reveals two sets of equations:
\begin{align}
\label{eq_l1}
   \frac{\partial L_{\text{CC}}}{\partial \overline{t}_{\nu;k}}&=\braket{\Phi_{\nu;k}|\hat{\overline{H}}|\Phi_{0}}=0 \quad k\le k_{\text{max}},\\
   \label{eq_l2}
   \frac{\partial L_{\text{CC}}}{\partial {t}_{\nu;k}}&=\braket{\Phi_{0}|(1+\hat{\Lambda})[\hat{\overline{H}},\hat{\tau}_{\nu;k}]|\Phi_{0}}=0  \quad k\le k_{\text{max}},
\end{align}
where Eq.~\eqref{eq_l1} are the usual amplitudes equations defined previously in Eq.~\eqref{eqn_amplitudes}, and Eq.~\eqref{eq_l2} are the \textit{Lambda equations}, a linear set of equations for the multipliers.\cite{Adamowicz1984}

We pointed out in Sec.\ref{sec:level2.2} the existence  of a connection between the left-eigenvalue problem reported in Eq.~\eqref{projected-left} and the computation of properties. To see this, we first expand the commutator in Eq.~\eqref{eq_l2} and insert a resolution of identity on the right-hand side
\begin{align}
\label{split}
    \braket{\Phi_{L}|\hat{\overline{H}}|\Phi_{\nu;k}}&= \braket{\Phi_{L}|\hat{\tau}_{\nu;k}\hat{\overline{H}}|\Phi_{0}}\\
    &=\sum_{m\neq0}^{N}\sum_{\eta}\braket{\Phi_{L}|\hat{\tau}_{\nu;k}| \Phi_{\eta;m}} \braket{\Phi_{\eta;m} |\hat{\overline{H}}|\Phi_{0}}\\
    &=\braket{\Phi_{L}| \Phi_{\nu;k}} E_{\text{CC}}
\end{align}
We then recover Eq.~\eqref{projected-left} since only the term with $m=0$ will survive. We may conclude that the lambda equations are the left-eigenvalue equations of the similarity-transformed Hamiltonian projected onto the excitation manifold. It should be noted that these latter equations, for first-order properties, do not depend on the perturbations, so we only need to solve these equations once. Similarly to what was done previously, for the multipliers we may invoke the $(2n+2)$ rule: the knowledge of the $n$-th derivative of the multipliers is sufficient to determine the $(2n+2)$ derivative of the CC Lagrangian.\cite{HTJBook} 

A significant difference between Eq.~\eqref{eq_l1} and Eq.~\eqref{eq_l2} is that the latter are not fully connected, meaning they can consist of disconnected diagrams. To see this, we reformulate the left-hand side of Eq.~\eqref{split} as
\begin{align}
\braket{\Phi_{L}|\hat{\overline{H}}|\Phi_{\nu;k}}=\braket{\Phi_{0}|\hat{\overline{H}}|\Phi_{\nu;k}}+\braket{\Phi_{0}|[\hat{\Lambda}, \hat{\overline{H}}]|\Phi_{\nu;k}}+\braket{\Phi_{0}|\hat{\overline{H}}\hat{\Lambda} |\Phi_{\nu;k}}.
\end{align}
Inserting a resolution of identity in the final term, we obtain\cite{Shavitt_Bartlett_2009}

\begin{align}
    \begin{split}
    \label{eqn:disconnected}
    \frac{\partial L_{\text{CC}}}{\partial t_{\nu;k}} = 0 &= \braket{\Phi_{0} | \hat{\overline{H}} | \Phi_{\nu;k}} + \braket{\Phi_{0} | [\hat{\Lambda}, \hat{\overline{H}}] | \Phi_{\nu;k}} \\
    &+ \sum_{m\neq0}^{N}\sum_{\eta} \braket{\Phi_{0} | \hat{\overline{H}} | \Phi_{\eta;m}} \braket{\Phi_{\eta;m} | \hat{\Lambda} | \Phi_{\nu;k}}, \quad k\le k_{\text{max}}
    \end{split}
\end{align}
where the term with $m=0$ was cancelled by the right-hand side of Eq.~\eqref{split}. Clearly, the $\Lambda_{1}$-equations are fully connected since the last term in Eq. (\ref{eqn:disconnected}) is then equal to zero; however, starting from the $\Lambda_{2}$ equations onward, we begin to encounter disconnected terms. 
As we will see in Sec.~(\ref{lambda-density}), using a diagrammatic approach leads to a simple and intuitive rule for predicting the disconnected diagrams appearing in the lambda equations.


The disconnected nature of the lambda equations will not affect the size-extensivity of properties. In fact, once Eq.~\eqref{eq_l1} and Eq.~\eqref{eq_l2} are satisfied, first-order molecular properties can be computed by using the Hellmann-Feymann theorem for approximate wave functions
\begin{align}
\begin{split}
\label{propA}
    \frac{dL_{\text{CC}}}{d\varepsilon_{A}}\Bigg|_{\bm{\varepsilon}=\bm{0}}&=\braket{\Phi_{0}|(1+\hat{\Lambda})\hat{\overline{H}}_{A}|\Phi_{0}},
    \end{split}
\end{align}
for an arbitrary perturbation $\varepsilon_{\text{A}}$. We can now introduce a commutator in the right-hand side of Eq.~\eqref{propA}
\begin{align}
\begin{split}
    \braket{\Phi_{0}|(1+\hat{\Lambda})\hat{\overline{H}}_{A}|\Phi_{0}}
    &=\braket{\Phi_{0}|\hat{\overline{H}}_{A}+[\hat{\Lambda},\hat{\overline{H}}_{A}]|\Phi_{0}}+\underbrace{\braket{\Phi_{0}|\hat{\overline{H}}_{A}\hat{\Lambda}|\Phi_{0}}}_{=0},
    \end{split}
\end{align}
where the last term is strictly zero since it has $\hat{\Lambda}$ acting on $\ket{\Phi_{0}}$.  We therefore conclude that
\begin{equation}
    \frac{d{L}_{\text{CC}}}{d\varepsilon_{A}}\Bigg|_{\bm{\varepsilon}=\bm{0}}=\braket{\Phi_{0}|\hat{\overline{H}}_{A}+[\hat{\Lambda},\hat{\overline{H}}_{A}]|\Phi_{0}},
\end{equation}
which means that calculating first-order molecular properties requires only connected terms, as required for size-extensivity.\cite{Shavitt_Bartlett_2009} Defining the perturbation as 
\begin{equation}
    \hat{H}_{A}=\sum_{pq}h_{pq;A}\{a_{p}^{\dagger}a_{q}\},
\end{equation}
and  by introducing the one-body CC response density matrices
\begin{equation}
\label{1dm}
   {\gamma}_{pq}^{\text{CC}}=\braket{\Phi_{0}|(1+\hat{\Lambda})e^{-\hat{T}}\{a_{p}^{\dagger}a_{q}\}e^{\hat{T}}|\Phi_{0}},
\end{equation}
which are computed by using the converged amplitudes and multipliers, we can compute  first-order properties by taking the trace of the product between the density operator and the operator matrix element
\begin{equation}
\label{exp_value}
    \frac{dL_{\text{CC}}}{d\varepsilon_{A}}\Bigg|_{\bm{\varepsilon}=\bm{0}}=\sum_{pq}h_{pq;A}\tilde{{\gamma}}^{\text{CC}}_{pq}
\end{equation}
It should be noted, however, that the density matrix in Eq.~\eqref{1dm} is not Hermitian, and thus the symmetrized form of the one-body density matrix is used to compute the expectation value in Eq.~\eqref{exp_value}
\begin{equation}
   \tilde{\gamma}_{pq}^{\text{CC}}=\frac{1}{2}({\gamma}_{pq}^{\text{CC}}+{\gamma}_{qp}^{\text{*CC}})
\end{equation}
In the present work, we have implemented the lambda equations and density matrices in \text{tenpi},\cite{brandejs2024generatingcoupledclustercode} following again the diagrammatic scheme of Kállay and Surján.\cite{Kallay2003,Kallay2004} As pointed out in Sec. \ref{CC_chapter}, since we are restricting ourself up to CCSDTQ, also the respective Lambda operator will contain de-excitation operators of maximum degree four
\begin{align}
    \hat{\Lambda}&=\hat{\Lambda}_{1}+\hat{\Lambda}_{2}+\hat{\Lambda}_{3}+\hat{\Lambda}_{4}\\
    \hat{\Lambda}_{1}&=\sum_{ia}\overline{t}_{a}^{i}a_{i}^{\dagger}a_{a} \\
    \hat{\Lambda}_{2}&=\left(\frac{1}{2}\right)^{2}\sum_{ijab}\overline{t}_{ab}^{ij}a_{i}^{\dagger}a_{j}^{\dagger}a_{b}a_{a}\\
     \hat{\Lambda}_{3}&=\left(\frac{1}{3!}\right)^{2}\sum_{ijkabc}\overline{t}_{abc}^{ijk}a_{i}^{\dagger}a_{j}^{\dagger}a_{k}^{\dagger}a_{c}a_{b}a_{a}\\
      \hat{\Lambda}_{4}&=\left(\frac{1}{4!}\right)^{2}\sum_{ijklabcd}\overline{t}_{abcd}^{ijkl}a_{i}^{\dagger}a_{j}^{\dagger}a_{k}^{\dagger}a_{l}^{\dagger}a_{d}a_{c}a_{b}a_{a}.
\end{align}
\subsection{\label{sec:level2.3} Rovibrationally-averaged properties}
\label{rovib-prp}

Within the Born–Oppenheimer approximation, the rovibrational structure of a diatomic molecule can be determined by solving the Schrödinger equation for an effective Hamiltonian\cite{Dunham1932}
\begin{equation}
\label{eqn:1}
\left(-\frac{\hbar^2}{2\mu}\frac{d^{2}}{dr^{2}} + V(r) + \frac{\hbar^{2}J(J+1)}{2\mu r^{2}}\right)\psi(r) = E\psi(r),
\end{equation}
where \(\hbar\) is the reduced Planck constant, \(\mu\) is the reduced mass, \(r\) is the internuclear distance, \(V(r)\) is the potential energy, \(J\) is the rotational quantum number (\(J=0,1,2,\ldots\)), \(\psi(r)\) is the rovibrational wavefunction, and \(E\) is the total energy.

To compute rovibrationally averaged properties, it is convenient to define the vibrational and rotational contributions to the Hamiltonian:
\begin{align}
\label{vib}
H_{\text{vib}} &= -\frac{\hbar^2}{2\mu}\frac{d^{2}}{dr^{2}} + V(r),\\[1mm]
\label{rotor}
H_{\text{rot}} &= \frac{\hbar^{2}J(J+1)}{2\mu r^{2}}.
\end{align}

We expand the potential \(V(r)\) in Eq.~(\ref{vib}) and the centrifugal term in Eq.~(\ref{rotor}) in a Taylor series around the equilibrium bond length \(r_e\). Introducing the normal coordinate \(x = r - r_e\), we write:
\begin{align}
    H_{\text{vib}} &= H_{\text{vib}}^{(0)} + \sum_{k=1}^{n} H_{\text{vib}}^{(k)},\\
    H_{\text{rot}} &= H_{\text{rot}}^{(0)} + \sum_{k=1}^{n} H_{\text{rot}}^{(k)},
\end{align}
where the zero-th order Hamiltonian corresponds to the harmonic oscillator and the rigid rotor:
\begin{align}
H_{\text{vib}}^{(0)} &= -\frac{\hbar^2}{2\mu}\frac{d^{2}}{dx^{2}} + \frac{1}{2}V^{[2]}x^2+V^{[0]}\\
H_{\text{rot}}^{(0)} &= \frac{\hbar^2 J(J+1)}{2\mu r_e^2},
\end{align}
and the higher-order terms are given by:
\begin{align}
H_{\text{vib}}^{(k)} &= \frac{1}{(k+2)!} V^{[k+2]} x^{k+2}, \quad k \ge 1,\\
H_{\text{rot}}^{(k)} &= \frac{\hbar^2 J(J+1)}{2\mu r_e^2} (-1)^k (k+1)\left(\frac{x}{r_e}\right)^k.
\end{align}
The constant terms $V^{[0]}$ and $\frac{\hbar^2 J(J+1)}{2\mu r_e^2}$ are then absorbed into the energy. 
Let us now consider a generic property \(P\) which depends on the displacement \(x\), and expand it in a Taylor series around the equilibrium geometry:
\begin{equation}
\label{expansion-property}
  P_{\nu J}(x) = P^{[0]} + P^{[1]}\braket{x}_{\nu J} + \frac{1}{2}P^{[2]}\braket{x^{2}}_{\nu J} + \dots,
\end{equation}
where
\[
P^{[k]} = \left.\frac{d^k P}{dx^k}\right|_{x=0}.
\]
By truncating Eq.~(\ref{expansion-property}) after \( P^{[2]} \), treating the cubic anharmonicity and first-order rotational correction perturbatively, and using harmonic oscillator wavefunctions as the unperturbed states, we arrive at the following expression for the rovibrationally averaged property:
\begin{equation}
\label{eqn:rovib}
   P_{\nu J}=P^{[0]}+P^{[2]}\left(\frac{\hbar}{\mu \omega_{e}}\right)(\nu+1/2)+P^{[1]}\frac{\hbar^{2}J(J+1)}{\mu^{2}\omega_{e}^{2}r_{e}^{3}}-\frac{1}{2}P^{[1]}V^{[3]}\left(\frac{\hbar}{\mu\omega_{e}}\right)^{2}\left(\frac{\nu+1/2}{\hbar\omega_{e}}\right)
\end{equation}
where \(\omega_e \) is the harmonic vibrational frequency. This expression corresponds to that originally derived by Buckingham in the context of temperature-dependent chemical shifts in NMR spectra.\cite{Buckingham1961} For a more extensive discussion on the evaluation of rovibrationally averaged properties, we refer the reader to Ref.~\citenum{Jameson1991}.

In order to get rovibrationally-averaged properties, we implemented Eq.~\eqref{eqn:rovib} in the utility program \text{VIBCAL}, which is part of the \text{DIRAC} code.\cite{DIRAC2020}

\newpage
\section{\label{sec:level3}Implementation}

In this section we will focus on the derivation and implementation of the expression for the lambda equations and one-body density matrix.

The ExaCorr module in DIRAC\cite{Pototschnig2021} provides X2C relativistic CCD/CC2/CCSD/CCSD(T) methods, based on the math libraries TAL-SH\cite{TAL_SH} and ExaTENSOR\cite{Lyakh2019,Pototschnig2021} developed by Lyakh. Recently,  Brandejs and co-workers also implemented high-order CC energy methods, i.e CCSDT and CCSDTQ in ExaCorr.\cite{brandejs2024generatingcoupledclustercode} The equations and the optimized codes were obtained using the \text{tenpi} toolchain; since this work is related to the expectation value module, we will only briefly summarize how the CC equations are generated from \text{tenpi}. 
\subsection{{tenpi}: Tensor Programming Interface}
\label{tenpi}
The automated generation of coupled cluster (CC) equations is important in particular when dealing with high-excitation methods (such as triples and quadruples), as their manual derivation is error-prone. Furthermore, the generation of intermediates with respect to a cost function (e.g. FLOP count, memory or communication) is a crucial task for ensuring computational efficiency. The construction of these intermediates involves numerous tensor contractions, and the higher the excitation level of the method, the greater the number of tensor indices involved. 

\text{tenpi} is an open-source coupled cluster code generator for modern distributed-memory architectures, developed by Brandejs and co-workers.\cite{brandejs2024generatingcoupledclustercode,tenpiGitlab} It performs the global optimization of intermediates and serves as a Fortran code generator. Additionally, \text{tenpi} features a Python-based user interface and a visualization module. One of the goals of \text{tenpi} \textit{"is to separate the science from the computational platform by getting tensor developments under the hood"},\cite{brandejs2024generatingcoupledclustercode} thus facilitating the development of coupled-cluster methods by simply inputting basic equations.

\text{tenpi} is based on the diagrammatic scheme introduced by Kállay and Surján,\cite{Kállay2001}  where each diagram is represented as a sequence of 13 integers. The last integer designates one of the 13 possible diagrammatic representations of the normal-ordered Hamiltonian. As shown in Eq.~(\ref{bch}), the similarity-transformed Hamiltonian is naturally truncated at the fifth term, meaning it can be contracted with up to four different cluster operators. The remaining 12 integers are divided into four triplets, each representing a connection (i.e., a contraction in the language of Wick's theorem) between the Hamiltonian and the $\hat{T}$ operators. In each triplet, the first index represents the excitation level of the cluster operator, the second indicates the number of connections between the Hamiltonian and this cluster operator, and the third denotes the number of particle (virtual) lines used to connect them. This integer-string representation efficiently generates distinct connected diagrams and identifies equivalent ones, as two diagrams are equivalent up to a sign if their strings are identical. See Ref.~\citenum{Kállay2001} for the original and complete algorithm. 

\text{tenpi} then translates each 13-integer string into a directed graph where a nodes of the graph corresponds to a node of a CC diagram vertex
and an edges correspond to particle and hole lines, i.e. the graphs are the diagrams themselves.
The graph is stored as a list of edges, where one edge is determined by its starting and ending node. 
After this step, CC rules are applied (see, e.g., Refs.~\citenum{CrawfordTDaniel2000AItC, Shavitt_Bartlett_2009} for a detailed discussion of CC rules), generating a list of tensor contractions. The directed-graph representation allows for intuitive implementation and easy customization of rules for different variants of CC. Intermediates are subsequently designed and optimized using \text{OpMin},\cite{LAI2012412} based on floating point operations (FLOP) count as a cost function. 

Once \text{tenpi} completes its optimization, the correctness of the intermediates is validated using a brute-force Python script. For further details on \text{tenpi}, we strongly encourage consulting Ref~\citenum{brandejs2024generatingcoupledclustercode}. 

In principle, \text{tenpi} can generate CC equations for an arbitrary excitation level. However, from a practical perspective, it does not take advantage of spatial symmetry and index-permutation symmetry,\cite{B803704N} which significantly impacts computational efficiency. As a result, excitations beyond the quintuple level are currently disabled. Efforts are underway to implement the missing symmetry support.

\subsection{Generation of the lambda equations and density matrices}
\label{lambda-density}
 As discussed in Sec.~\ref{sec:level2.3}, computing CC expectation values requires the evaluation of the CC one-body density matrices, Eq.~(\ref{1dm}), which are constructed from the converged amplitudes (Eq.~(\ref{eq_l1})) and multipliers (Eq.~(\ref{eq_l2})). While the algorithm for generating amplitudes is already well-established,\cite{brandejs2024generatingcoupledclustercode} additional considerations are required to extend the same approach to the multipliers.  

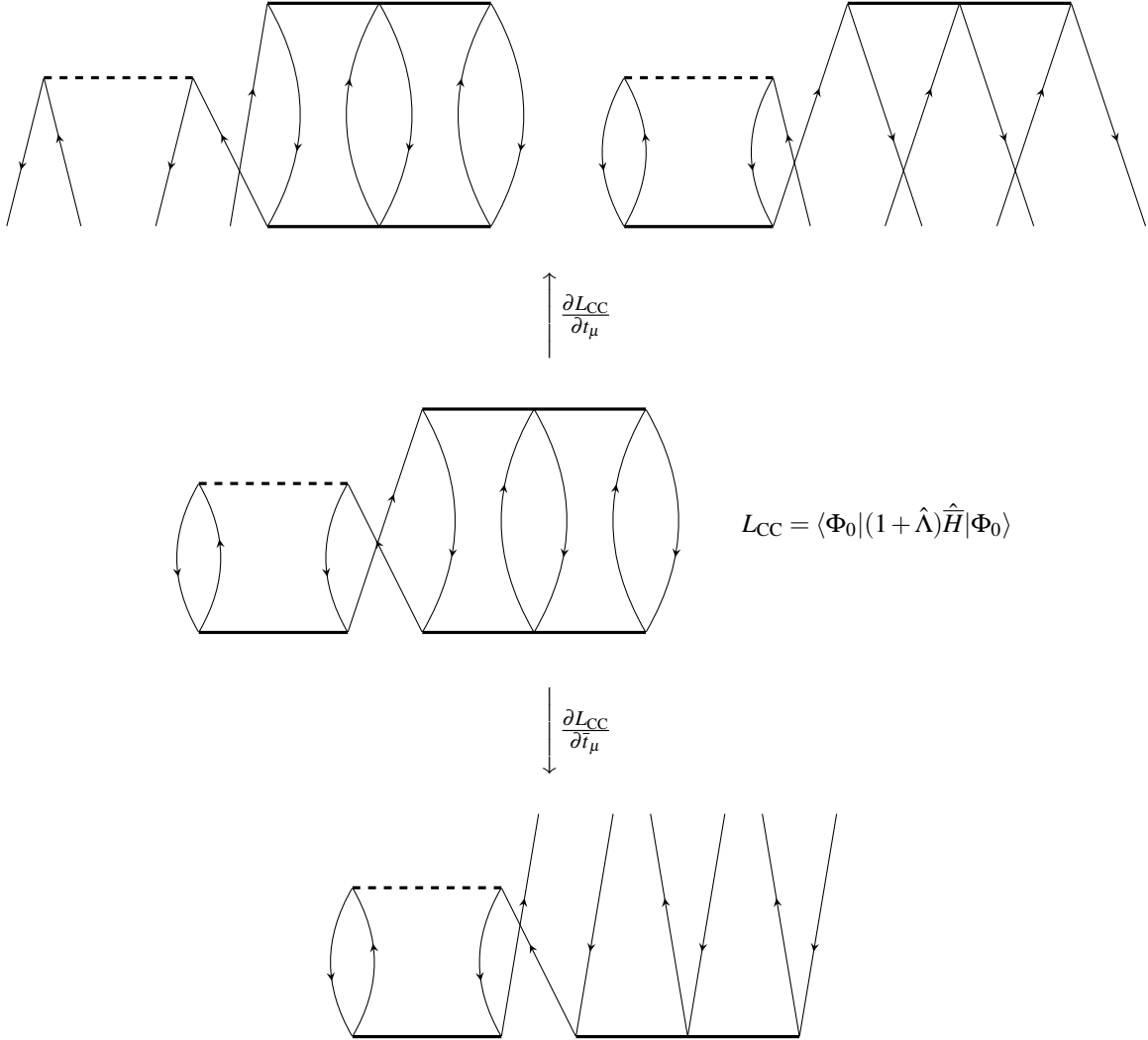
\begin{figure}[h!]
\centering
\begin{center}
\diagsav[LAMBDA-T2]
{\bdiag[2.0]
\dWs{W_1_1}{W_1_2}
\dmoveTd{6}
\dmoveT{6}
\dTs[]{3}{T3_}
\dTds[]{3}{td_}
\dline[]{W_1_1}{W_1_1vd1}
\dline[]{W_1_1vd2}{W_1_1}
\dline[]{W_1_2}{W_1_2vd1}
\dline[]{W_1_2vd2}{td_1}
\dline[]{T3_1}{W_1_2}
\dcurver[]{td_1}{T3_1}
\dcurver[]{T3_2}{td_2}
\dcurver[]{td_2}{T3_2}
\dcurver[]{T3_3}{td_3}
\dcurver[]{td_3}{T3_3}
\ediag} 
\diagsav[LAMBDA-T3]
{\bdiag[2.0]
\dWs{W_1_1}{W_1_2}
\dmoveTd{6}
\dTs[]{2}{T2_}
\dTds[]{3}{td_}
\dcurve[]{W_1_1}{T2_1}
\dcurve[]{T2_1}{W_1_1}
\dcurve[]{W_1_2}{T2_2}
\dline[]{T2_2}{td_1}
\dline[]{W_1_2vd2}{W_1_2}
\dline[]{td_1}{td_1v2}
\dline[]{td_2v1}{td_2}
\dline[]{td_2}{td_2v2}
\dline[]{td_3v1}{td_3}
\dline[]{td_3}{td_3v2}
\ediag}
\vskip 5mm
$\Bigg\uparrow\frac{\partial L_{\text{CC}}}{\partial {t}_{\mu}}$\\
\vskip 5mm
\begin{tikzpicture}
  \node (diag) at (0,0) {
    \diagsav[LAP-org]{
      \bdiag[2.0]
        \dWs{W_1_1}{W_1_2}
        \dmoveTd{6}
        \dTs[]{2}{T2_}
        \dTs[]{3}{T3_}
        \dTds[]{3}{td_}
        \dcurve[]{W_1_1}{T2_1}
        \dcurve[]{T2_1}{W_1_1}
        \dcurve[]{W_1_2}{T2_2}
        \dline[]{T2_2}{td_1}
        \dline[]{T3_1}{W_1_2}
        \dcurver[]{td_1}{T3_1}
        \dcurver[]{T3_2}{td_2}
        \dcurver[]{td_2}{T3_2}
        \dcurver[]{T3_3}{td_3}
        \dcurver[]{td_3}{T3_3}
      \ediag
    }
  };

  \node at (6,0) {\footnotesize $L_{\text{CC}} = \braket{\Phi_{0}|(1+\hat{\Lambda})\hat{\overline{H}}|\Phi_{0}}$};
\end{tikzpicture}
\vskip 5mm
$\Bigg\downarrow\frac{\partial L_{\text{CC}}}{\partial \overline{t}_{\mu}}$
\vskip 5mm
\diagsav[AMP-org]
{\bdiag[2.0]
\dWs{W_1_1}{W_1_2}
\dTs[]{2}{T2_}
\dTs[]{3}{T3_}
\dcurve[]{W_1_1}{T2_1}
\dcurve[]{T2_1}{W_1_1}
\dcurve[]{W_1_2}{T2_2}
\dline[]{T2_2}{T2_2v2}
\dline[]{T3_1}{W_1_2}
\dline[]{T3_1v2}{T3_1}
\dline[]{T3_2}{T3_2v1}
\dline[]{T3_2v2}{T3_2}
\dline[]{T3_3}{T3_3v1}
\dline[]{T3_3v2}{T3_3}
\ediag}
\end{center}
\caption{Diagrammatic structure of the coupled‐cluster Lagrangian: the central diagram is obtained by capping all open lines in the amplitude equations with the de‐excitation operator \(\hat{\Lambda}\); the downward arrow indicates differentiation with respect to the multipliers \(\overline{t}_{\mu}\), and the upward arrow indicates differentiation with respect to the cluster amplitudes \(t_{\mu}\).}
\label{diagrams-capping}
\end{figure}

The original algorithm by K\'allay and co-workers,\cite{Kallay2003}is based on the following observations: the generation of the CC Lagrangian  is based on capping the original amplitude equations by a de-excitation operator \( \hat{\Lambda} \) from above and therefore closing all the open upward lines, as exemplified by the central diagram in Figure (\ref{diagrams-capping}) going from the bottom diagram to the middle one. Taking the derivative of the Lagrangian with respect to the particular $\hat{\Lambda}$-amplitude implies going the other way, hence removing the $\hat{\Lambda}$ interaction line. From this we learn that differentiating with respect to a particular $\hat{T}$-amplitude correspond to sequentially removing the corresponding $\hat{T}$ interaction lines. In Figure (\ref{diagrams-capping}), starting from the Lagrangian element, this generates the two top diagrams. This frees up a slot in Kállay's string, which is then filled with an integer referring to the  \( \hat{\Lambda} \)-operator. For full details of these operations and their diagrammatic representation, please refer to Ref.\citenum{Kallay2003}.

\normalsize
\newpage
\begin{figure}[h!]
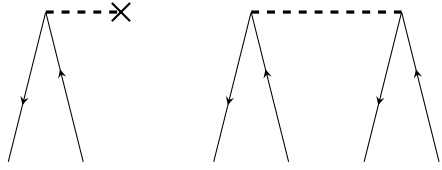

\centering
\begin{center}
\diagsav[F-1]
{\bdiag[2.0]
\dFs{F}
    \dmoveTd{6}
    \dmoveT{6}
    \dline[]{F}{Fvd1}
    \dline[]{Fvd2}{F}
\ediag}\qquad
\diagsav[W-2]
{\bdiag[2.0]
\dWs{W_1_1}{W_1_2}
\dmoveTd{6}
\dmoveT{6}
\dline[]{W_1_1}{W_1_1vd1}
\dline[]{W_1_1vd2}{W_1_1}
\dline[]{W_1_2}{W_1_2vd1}
\dline[]{W_1_2vd2}{W_1_2}
\ediag}
\end{center}
\caption{Elements of the Hamiltonian contributing to disconnected diagrams in the \( \hat{\Lambda} \)-equations.}
\label{F-W diagrams}
\end{figure}
This diagrammatic framework allows us to explicitly show from which amplitude equations the disconnected diagrams in the \(\hat{\Lambda}\) equations originate.  To do so, we recall that the Hamiltonian \(\hat{H}_{N}\) must be connected to all possible cluster operators \(\hat{T}\) (up to four clusters at most). Disconnected diagrams can only occur when the 
\( \hat{\Lambda} \)-operator is not directly connected to the Hamiltonian, which limits the possible elements to those given in Figure \ref{F-W diagrams}, having \textit{no} upward lines and an \textit{even} number of downward lines. In order to be disconnected upon removal of a \(\hat{T}\)-interaction line, the \(\hat{\Lambda}\) has to be connected to that \(\hat{T}\) only, hence by an \textit{even} number of connections, while the \(\hat{T}\) at the same time must be connected to the Hamiltonian. This excludes disconnected diagrams from the \(\hat{\Lambda}_{1}\)-equations, as we concluded by more elaborate means in Section \ref{sec:level2.3}. More generally, we also learn that at least one downward external line must emanate from the Hamiltonian, as indicated by the commutator appearing in Eq.~\eqref{eq_l2}.

 In our approach, we generate the lambda equations directly from matrix elements produced by Eq.~\eqref{eq_l2} that incorporate both the similarity–transformed Hamiltonian and the $\hat{\Lambda}$ operator. Although we also close the amplitude diagrams with the $\hat{\Lambda}$ operator from above, our integer string encoding remains unmodified. In our method the lambda diagrams are obtained by simply appending $\hat{\Lambda}$ to the amplitude diagrams, thereby preserving the original structure and streamlining the generation process. Once these connections are established, \text{tenpi} processes the diagrams as described in Sec.~\ref{tenpi}, with subsequent factorization and reordering to ensure efficient contraction of intermediates.

For the generation of the one-body density matrices, we follow the idea proposed by Kállay,\cite{Kallay2003} noting that Eq.~(\ref{1dm}) fundamentally involves, in addition to the $\hat{\Lambda}$ operator and $\hat{\overline{H}}$, a one-electron operator $\{a_{p}^{\dagger}a_{q}\}$. Therefore, to represent this one-electron operator, in \text{tenpi} we replace $\{a_{p}^{\dagger}a_{q}\}$ by $\hat{F}=\sum_{pq}f_{pq}\{a_{p}^{\dagger}a_{q}\}$, and once the diagrams are generated, we remove the matrix element \( f_{pq} \). Again, once the diagrams and the algebraic expressions are generated, they are processed by \text{tenpi} as explained in Sec.\ref{tenpi}.

The generation of the lambda equations involving excitations beyond doubles (i.e., triples and quadruples) took longer than expected due to several bugs. As widely documented in the literature,\cite{CrawfordTDaniel2000AItC,BartlettRev20007,Shavitt_Bartlett_2009} each pair of equivalent lines (i.e., lines beginning at the same horizontal interaction line and ending at the same interaction line) contributes a factor of \(1/2\) to the algebraic representation of a diagram. This rule accounts for the fact that, both in the amplitude equations (of any order) and in the lambda equations (up to double excitations), at most \textit{pairs} of equivalent lines are encountered. However, in the specific case of the lambda equations (as well as for density matrices), the inclusion of triple and higher excitations requires an extension of the equivalent line rule. In these cases, groups of three, four, or more equivalent lines can appear, as seen for example in the upper left diagram of Figure (\ref{diagrams-capping}). Consequently, if \(n\) is the number of equivalent lines, a factor of \(1/n!\) must be applied to the algebraic representation of the diagram. This rule is also mentioned in a footnote of Ref.\citenum{CrawfordTDaniel2000AItC}. 



During CCSDT/Q intermediate optimization, we found that Opmin’s integer-only \texttt{math.gcd}\cite{python_math_gcd} python function caused instability during factorization when non-integer floating-point prefactors were present. We implement the following sequence to fix this: i) compute the least common multiple (LCM) across all prefactor denominators, ii) scale each tensor contraction by that LCM to yield integer prefactors, iii) run Opmin, iv) divide the results by the LCM to restore the original scale.

\section{\label{sec:level4}Computational Protocol}
All calculations were performed with development versions of
the DIRAC\cite{DIRAC2020} and the TAL-SH library. Details on the
particular revisions used in the calculations described below are given in the respective output files that are provided within
a separate repository.\cite{ReplicationData} In this study, we mainly use the exact two-component (X2C) Hamiltonian with molecular mean-field correction.\cite{Sikkema2009} The Gaussian model of nuclei has been employed for all the molecules  using the parameters of Ref.~\citenum{Visscher:atomic}. We performed all calculations including the Gaunt interaction at the SCF level, as well as the $\braket{\text{SS}|\text{SS}}$ class of two-electron integrals. 

Our first goal was to verify the correctness of the new high-order CC expectation values. To do so, we first implemented the CCSD model using the autogenerated code from \text{tenpi} and compared it with the hand-coded implementation already present in \text{ExaCorr}.\cite{Pototschnig2021} Once this validation was completed, to further assess the correctness of the CCSDT and CCSDTQ expectation values, we compared the results of the new implementations using TAL-SH to those obtained with the MRCC quantum chemistry package,\cite{kallay2020mrcc} using a non-relativistic Hamiltonian.

To define the active space, we employ the notation AS(k,n), where k represents the number of electrons and n the number of Kramers pairs involved in the coupled cluster calculation (occupied plus virtual). 

For the non-relativistic calculations, we have considered the computation of the dipole moment of LiH in a 6-31G basis\cite{Ditchfield1971,Dill1975} at an internuclear distance of 3.015 $a_{0}$.\cite{NIST_Diatomic_Spectral_Database}. Subsequently, a more relevant application of our high-order CC expectation values code is the determination of the nuclear quadrupole moment (NQM) of $^{7}$Li and $^{27}$Al from molecular data. To determine $Q(\ce{^7Li})$, we considered three molecules: LiH, LiF, and LiCl. The experimental equilibrium bond distances of these molecules, 3.015 $a_{0}$, 2.9553 $a_{0}$, and 3.81852 $a_{0}$, respectively, were taken from Ref.\citenum{URBAN1990157}. For all the lithium compounds we took the dyall.ae4z\cite{Dyall2016} basis set, where the latter was augmented with a tight p-exponent with value \(435.417100 a_{0}^{-2}\) with a tight d-exponent with value \(62.168460 a_{0}^{-2}\), as explained in Sec.\ref{5.B.1}. For all three molecules, CCSD calculations were performed correlating all electrons. The active space was defined by including the full set of available virtual orbitals.

 For the AlY (Y = H, F, Cl, Br) calculations, we used the equilibrium distances 3.1129~$a_0$, 3.1263~$a_0$, 4.0241~$a_0$ and 4.3372~$a_0$. For the CCSD calculations, we employed the basis dyall.aae4z,\cite{Dyall2006} which was augmented by a p-exponent with value 24750.350000 $a_{0}^{-2}$, a d-exponent with value 1233.029836 $a_{0}^{-2}$ and a f-exponent with value 548.276242 $a_{0}^{-2}$ as explained in Sec.\ref{5.B.1}. 
 
 For all the systems, to compute the triple and quadruple corrections, we used a slightly smaller basis, {dyall.v3z}.\cite{Dyall2016,Dyall2006}  
 
 \subsubsection{Basis set and Active space}
\label{5.B.1}

Since basis sets are generally optimized for energy rather than properties, and because we start from basis sets already optimized for correlated calculations, we investigated the convergence of our basis set with respect to the EFG at the SCF level. 

Tables~\ref{tab:nqcc_relative_error}--\ref{tab:aluminum_spectroscopy} report the experimental values of the nuclear quadrupole coupling constants, along with their associated uncertainties. To ensure that the computational error does not exceed the relative uncertainties, we require that the relative change in \( q \) due to basis set augmentation remains below the experimental relative uncertainty associated to the NQCCs. For the lithium systems, we found that the smallest relative uncertainty is associated to \ce{^{7}Li^{1}H}, which is 0.0072\%, and therefore we require that, in order to ensure that the accuracy of the extracted quadrupole moment \(Q\) depends only on the experimental error, the accuracy of the EFG must be determined with a relative uncertainty lower than 0.0072\%, whereas for the aluminum systems, we found that the smallest relative uncertainty is associated with  $^{27}$Al$^{35}$Cl, which is 0.0089\%.  

\begin{table}[h]
    \centering
    \begin{tabular}{lcc}
        \hline
        Molecule & NQCC (MHz) & Relative Uncertainty (\%) \\
        \hline
        \ce{^{7}Li^{1}H} ($\nu=0$, $J=1$) & $0.346750 \pm 0.000025$\cite{{Rothstein1969}} & 0.0072\\
        \ce{^{7}Li^{1}H} ($\nu=1$, $J=1$) & $0.332 \pm 0.005$\cite{{Rothstein1969}}        & 0.0015\\
        \ce{^{7}Li^{2}H} ($\nu=0$)        & $0.349 \pm 0.001$\cite{{Rothstein1969}}        & 0.2865\\
        \ce{^{7}Li^{19}F} ($\nu=0$)       & $0.4156 \pm 0.0004$\cite{{Rothstein1969}}       & 0.0963\\
        \ce{^{7}Li^{19}F} ($\nu=1$)       & $0.4061 \pm 0.0006$\cite{{Rothstein1969}}       & 0.1477\\
        \ce{^{7}Li^{19}F} ($\nu=2$)       & $0.3965 \pm 0.0008$\cite{{Rothstein1969}}       & 0.2016\\
        \ce{^{7}Li^{35}Cl} ($\nu=0$)      & $0.24993 \pm 0.00050$\cite{{Rothstein1969}}     & 0.2000\\
        \ce{^{7}Li^{35}Cl} ($\nu=1$)      & $0.2446 \pm 0.0020$\cite{{Rothstein1969}}                             & 0.8179\\
        \hline
    \end{tabular}
    \caption{NQCC and relative error with respect to uncertainty for the lithium systems.}
    \label{tab:nqcc_relative_error}
\end{table}

\begin{table}[h]
    \centering
    \begin{tabular}{lcc}
        \toprule
        Molecule & NQCC (MHz) & Relative Uncertainty (\%) \\
        \midrule
        \(^{27}\)Al\(^{1}\)H & \(-48.59 \pm 0.70\)\cite{9d88811980d04af9a5ca70a30b620871} & 1.44 \\
        \(^{27}\)Al\(^{2}\)H & \(-48.48 \pm 0.88\)\cite{9d88811980d04af9a5ca70a30b620871} & 1.815 \\
        $^{27}$Al$^{19}$F, $R_e$ & $-37.75 \pm 0.08$\cite{Lovas1974_JPCRD} & 0.212 \\
        $^{27}$Al$^{19}$F ($v = 0$) & $-37.53 \pm 0.12$\cite{Lovas1974_JPCRD} & 0.320 \\
        $^{27}$Al$^{35}$Cl ($v = 0$) & $-30.4081 \pm 0.0027$\cite{Lovas1974_JPCRD} & 0.0089 \\
        $^{27}$Al$^{79}$Br ($v = 0$) & $-28.0059 \pm 0.0035$\cite{1985OptSp..59..699A} & 0.0125 \\
        \bottomrule
    \end{tabular}
    \caption{NQCC and relative error with respect to uncertainty for the aluminum systems.}
    \label{tab:aluminum_spectroscopy}
\end{table}


We employ the following protocol: We first add exponents in an even‐tempered manner until satisfactory convergence of the EFG is achieved. However, this typically produces a very large basis set that is too costly for high‐order CC calculations; therefore, we then perform a line search to optimize a single exponent against the reference value obtained with the large basis set. This procedure ensures an optimal compromise between computational cost and EFG accuracy. 

The \(s\) exponents were omitted because projection analyses\cite{Dubillard_JCP2006,Fabbro_JPCA_2025} show that \(s\) orbitals, being spherically symmetric, contribute negligibly to the EFG. Furthermore, our previous work\cite{Fabbro_JPCA_2025} demonstrated near‐complete cancellation between electronic and nuclear contributions from surrounding nuclei. The dominant effect is orbital polarization by ligands, so we focus on the most polarized systems, LiF and AlF.

Starting from dyall.ae4z (19s10p5d3f) for Li and dyall.aae4z (25s15p10d7f5g) for Al, we added one p‐exponent (435.417100 \(a_0^{-2}\)) and one d‐exponent (62.168460 \(a_0^{-2}\)) to Li—yielding a 19s11p6d3fset—and one p‐exponent (24750.350000 \(a_0^{-2}\)), one d‐exponent (1233.029836 \(a_0^{-2}\)), and one f‐exponent (548.276242 \(a_0^{-2}\)) to Al—yielding a 25s16p11d8f5g set. The addition of d-type exponents led to slow EFG convergence for both systems—unexpectedly also in Li, which lacks occupied d orbitals—due to the EFG operator coupling functions differing by two units of angular momentum. Further details can be found in the Supporting Information.



Since the current high‐order expectation‐values module runs entirely on a single compute node, it is constrained by the available RAM on that node; hence, we can not include a very large number of virtual spinors without exceeding the memory capacity of the Olympe machine that we are currently using.\cite{top500} One of the future goals we are already working on is the transition to multiple-nodes using the Cyclops library.\cite{solomonik2014massively} Moreover, \text{ExaCorr} currently does not exploit point-group symmetry and index permutation symmetry of tensors.\cite{Pototschnig2021} So, the major disadvantage is that the active space (AS) is relatively small. 

Therefore, we have used a composite approach in order to determine the EFG at the lithium and aluminum nuclear positions. 
An exemplary instance of such a procedure at the CC level was proposed recently by Skripnikov in the context of determining the nuclear magnetic and electric quadrupole moments of polonium isotopes,\cite{Skripnikov_PhysRevC.109.024315} as well also the determination of the nuclear electric quadrupole moment of isotopes $^{33}$S and $^{35}$S by Stopkowicz and Gauss.\cite{PhysRevA.90.022507} 
The strategy adopted was to obtain an accurate EFG value evaluated at the CCSD level using the RELCCSD\cite{Visscher1996,DIRAC2020,Shee2006} module (which benefits from  point-group symmetry) with a large active space and a sufficiently extended basis set, and then subsequently calculate the corrections arising from triple and quadruple excitations. We have estimated the triple and quadrupole correction by performing calculations on a smaller basis set, the \text{dyall.v3z}.\cite{Dyall2016} Therefore, the contribution of triple excitations can then be estimated as
\begin{equation}
    T = \braket{q}_{\text{CCSDT}}^{\text{dyall.v3z}} - \braket{q}_{\text{CCSD}}^{\text{dyall.v3z}},
\end{equation}
using the same active space for the CCSD and CCSDT calculations, as well for the quadruples
\begin{equation}
    Q = \braket{q}_{\text{CCSDTQ}}^{\text{dyall.v3z}} - \braket{q}_{\text{CCSDT}}^{\text{dyall.v3z}},
\end{equation}
where \( \braket{q} \) represents the electric field gradient expectation value. This approximate contribution can then be added to the value computed with a \textit{large} basis set, determinate as explained above
\begin{equation}
    \braket{\Omega}_{\text{Q(T(CCSD))}} =\braket{\Omega}_{\text{CCSD}}^{\text{opt-basis}} + T+Q,
\end{equation}
where \text{opt-basis} is the optimized basis discussed before, i.e \(\text{19s11p6d3f}\) for Li and \text{25s16p11d8f5g} for Al.

\begin{figure}
    \centering
    \includegraphics[width=1.0\linewidth]{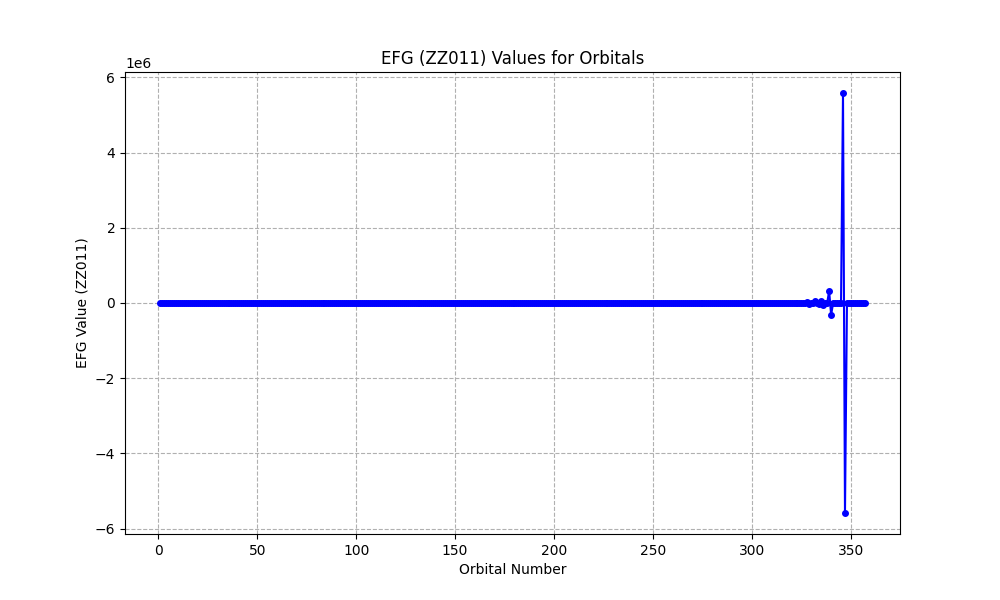}
    \caption{Contribution to the EFG from the HF orbitals in AlF molecule.}
    \label{AlF-EFG-orbitals-HF}
\end{figure}

At the CCSD level, we fully correlated all electrons in LiX (X = H, F, Cl) systems owing to their small size. For AlY (Y=H, F, Cl, Br), however, memory constraints forced us to apply an energy cutoff when selecting virtual orbitals—a critical choice for accurate EFGs. The energy cutoff should be chosen so as not to completely eliminate correlation functions. Pernpointner and Visscher showed that excluding orbitals above 4.5 $E_{h}$ (AlF, AlCl) and 4.4 $E_{h}$ (AlBr) avoids sharp EFG oscillations.\cite{Pernpointner2001_JCP} A contribution to such unstability is as follows: Figure \ref{AlF-EFG-orbitals-HF}, we report the contributions of the Hartree–Fock orbitals to the electric field gradient in AlF. A pronounced oscillation of the order of \(10^{6}\,E_{h}/e a_{0}^{2}\), attributable primarily to orbitals 346 and 347, is observed, which swamps all the other contributions. Projection analysis\cite{Dubillard_JCP2006} reveals that these orbitals correspond to aluminum $p_{3/2}$-type spinors, with $|m_j| =  1/2$ and $ 3/2$. Notably, orbital 346 contributes $5562233.8\,E_h/e a_0^2$, while orbital 347 contributes $-5562392.7\,E_h/e a_0^2$. In the free atom, these contributions should cancel,\cite{Fabbro_JPCA_2025} but in the molecule their net contribution to the EFG is $-158.9\,E_h/e a_0^2$. The effect is therefor due to polarization of these core-like virtual orbitals in the molecule. In fact, orbital 345, identified as the partner $p_{1/2}$ spinor, contributes $156.3\,E_h/e a_0^2$, which almost cancels out the residual contributions from orbitals 346 and 347, but not entirely. It follows that noise may be introduced in the correlated calculations if the energy threshold only partially includes an atom-like shell.

For AlF, we observed that the EFG stabilizes once a threshold of 45\,$E_{h}$ is reached. Extending the energy cutoff beyond this threshold does not result in any significant variation in the EFG, nor does it affect the final quadrupole moment. We retained this cutoff also for AlCl and AlBr to keep the calculations computationally feasible without compromising on the correlation of inner-shell electrons.

 Moving to higher-order coupled-cluster methods, tighter restrictions on the number of correlated electrons and orbitals were necessary to keep the computations feasible. Table~\ref{tab:comparison-li-al} summarizes the corresponding number of correlated electrons and spinors for each system and method. The notation $(n,k)$ indicates that $n$ electrons and $k$ spinors (occupied+virtuals) were correlated in the calculations.
 
\begin{table}[h]
    \centering
    \begin{tabular}{lccccccc}
        & \multicolumn{3}{c}{LiX} & \multicolumn{4}{c}{AlY} \\
        & LiH & LiF & LiCl & AlH & AlF & AlCl & AlBr \\
        \hline
        CCSD   & (2, 300) & (6, 412) & (10, 566) & (14, 398) & (22, 474) & (30, 560) & (48, 584) \\
        CCSDT  & (2, 78)  & (3, 68)  & (3, 68)   & (3, 68)   & (3, 68)   & (3, 68)   & (3, 68) \\
        CCSDTQ & (2, 18)  & (3, 18)  & (3, 18)   & (3, 18)   & (3, 18)   & (3, 18)   & (3, 18) \\
    \end{tabular}
    \caption{For each molecule and each method, the number of correlated electrons and the number of spinors used in the active space are reported.}
    \label{tab:comparison-li-al}
\end{table}

\section{\label{sec:level5}Sample Applications}

\subsection{Correctness: dipole moment of LiH}
To confirm the correctness of our high-order coupled-cluster expectation value code, we compared the ground-state coupled cluster dipole moment of LiH from \text{tenpi} (this work) with those from the MRCC package\cite{kallay2020mrcc,mrcc2024,Kallay2003} and, where feasible, with the DIRAC code (using the hand-coded one).\cite{Pototschnig2021} 
From Table \ref{tab:convergence_total_dipole_moment} we see that, at SCF level, dipole moments computed with DIRAC and MRCC are virtually identical and indicate that both codes yield equivalent SCF-level results, thus providing a solid baseline for subsequent correlated methods.
For the CCSD method, the dipole moments computed with DIRAC using both the \text{tenpi} autogenerated code and the hand-coded (hc) implementation are exactly the same. The MRCC package produces a minute difference (on the order of \(10^{-8}\)), which is negligible and demonstrates that our autogenerated code is fully consistent with the handwritten implementation as well as with an independent reference from MRCC. 
At higher levels of theory, i.e., CCSDT (including triple excitations) and CCSDTQ (including quadruple excitations), the dipole moments from DIRAC (\text{tenpi}) and MRCC remain in excellent agreement in the convergence threshold limit.

\begin{table}[H]
    \centering
    \begin{tabular}{|l|l|c|c|}
        \hline
        \textbf{method} & \textbf{code} & \textbf{convergence} & \textbf{Total dipole moment [$E_{h}/ea_{0}$]}\\
 SCF& DIRAC & &-2.33268232\\
 SCF& MRCC& &-2.33268233\\ \hline
        CCSD & DIRAC \text{tenpi} & \num{0.6E-09}& -2.16553800 \\
 CCSD& DIRAC hc& \num{0.6E-09}&-2.16553800 \\
        CCSD & MRCC & \num{1.3E-10}& -2.16553802\\ 
        CCSDT & DIRAC \text{tenpi} & \num{0.9E-09}& -2.16528064\\ 
        CCSDT & MRCC & \num{1.8E-10}& -2.16528066\\ 
        CCSDTQ & DIRAC \text{tenpi} & \num{0.6E-09}& -2.16527943\\
        CCSDTQ & MRCC & \num{2.1E-10}& -2.16527945\\ \hline
    \end{tabular}
    \caption{Convergence of the final error shown in the run and Total dipole moment for Different Methods and Codes for LiH.  hc=hand-coded.}
    \label{tab:convergence_total_dipole_moment}
\end{table}

\newpage
\subsection{Nuclear Electric Quadrupole Moments of  $^{7}$Li}

Pioneering values of the EFG for lithium were obtained using the restricted Hartree-Fock (RHF)\cite{Kahalas1963,Cade1967} and Configuration-Interaction (CI) calculations.\cite{Browne1964,Bender1966} A significant improvement in the calculation of the EFG at the lithium nucleus was provided by Green,\cite{Green1971} who used a CI wavefunction including 200 determinants. However, as pointed out shortly thereafter by Sundholm and co-workers,\cite{Sundholm1984} the value reported by Green might not have been sufficiently accurate due to basis set incompleteness. Moreover, Green's calculations relied on the Hellmann-Feynman theorem\cite{Hellmann1937,Feynman1938} applied to a limited CI function, potentially introducing inaccuracies in the correlation correction to \(q\). 

To address these issues, Sundholm and co-workers used a hybrid approach that combined fully numerical solutions of the HF equations with discrete basis set calculations for correlation corrections. 
Additionally, Sundholm and co-workers ensured that their wavefunctions satisfied the Hellmann-Feynman theorem by employing a full CI approach within the chosen active space, leading to more accurate correlation corrections and incorporating rovibrational effects. Their work provided a benchmark for the EFG, refining \(Q(^{7}\text{Li})\) to \(-0.0406 \, \text{b}\).\cite{Sundholm1984}

Shortly thereafter, Diercksen and co-workers extended the analysis to lithium fluoride (LiF).\cite{Diercksen1988} 
Their findings reinforced the molecular value of \(Q(^{7}\text{Li})\) as \(-0.04055(80) \, \text{b}\), consistent with results obtained from LiH,\cite{Sundholm1984} subsequently confirmed by Urban and Sadlej, also including the LiCl molecule in the analysis, \cite{Karna1990,URBAN1990157} getting $Q(\ce{^7Li})$=0.0406 b. This value was later adopted as the reference by Pyykkö.\cite{Pyykko1992}

Recently, Guan and co-workers,~\cite{Guan:2024oyp} using an optical Ramsey technique combined with bound-state quantum electrodynamics theory, determined the quadrupole moment of \ce{^7Li} as $Q(\ce{^7 Li})=$ -0.0386(5) b, which clearly deviates from the currently accepted one.\cite{URBAN1990157}

On our side, we aim to shed light on this discrepancy through the methodology previously discussed, which combines state-of-the-art relativistic coupled-cluster calculations with an accurate treatment of rovibrational effects.

We have reported in Table~\ref{tab:li_efg_contributions} the different values of the EFG evaluated ad the lithium position for the three different systems LiH, LiF and LiCl.  Table~\ref{tab:li_efg_contributions} reveal that inclusion of electron correlation through CCSD introduces modest corrections that slightly adjust the HF values; the additional full triples corrections further refine the results, although the full quadruple corrections are negligible, indicating that higher-order correlation effects beyond triples are minimal. 

\begin{table}[h!]
    \centering
    \begin{tabular}{lcccccc}
        \toprule
        & \multicolumn{2}{c}{LiH} & \multicolumn{2}{c}{LiF} & \multicolumn{2}{c}{LiCl} \\
        \cmidrule(lr){2-3} \cmidrule(lr){4-5} \cmidrule(lr){6-7}
        & EFG & $Q$ & EFG & $Q$ & EFG & $Q$ \\
        \midrule
        HF & -0.039726 & -0.037148 & -0.049030 & -0.036075 & -0.027471 &-0.038721  \\
        + CCSD & 0.001020  &-0.002578  &0.002742  & -0.002137 &  -0.000316& 0.000440 \\
        + correction for full triples &0.000051  & 0.001549 &0.000279  & -0.000232 & 0.000275 &  -0.000382\\
        + corrections for full quadruples &0.000000  & 0.000000 & 0.000000 &  0.000000&0.000000  &0.000000  \\
        final value & -0.038655 & -0.038177 & -0.046009 & -0.038444 & -0.027512 & -0.038663 \\
        \bottomrule
    \end{tabular}
    \caption{Individual contributions to the EFG ($E_{h}/ea_{0}^{2}$) at the Li position for the systems LiX (X = H, F, Cl) and the derived values for the $^{7}$Li quadrupole moment ($Q$, in mb).}
    \label{tab:li_efg_contributions}
\end{table}
To better relate our theoretical EFG results with the experimental nuclear quadrupole coupling constants, we took into account the rovibrational contributions, as detailed in Sec.~\ref{sec:level2.3}. In this context, the rovibrationally averaged EFGs were computed using the utility program \text{VIBCAL} included in the \text{DIRAC} program package. In order to compute these quantities, for each molecule a potential energy surface (PES) and a corresponding property surface (PS) were generated by sampling multiple points along the molecular coordinates at the CCSD+T+Q level of theory. The individual electronic contributions are summarized in Table~\ref{tab:li_efg_contributions}, while the final rovibrationally averaged values are reported in Table~\ref{tab:li_efg}. The vibrational corrections, although modest, lead to non-negligible adjustments of the final electronic values. For example, in the case of LiH (for $\nu=0$, $J=1$), the vibrationally averaged EFG is $-0.037682\,E_{h}/ea_{0}^{2}$, compared to the electronic final value of $-0.038655\,E_{h}/ea_{0}^{2}$, corresponding to a reduction of approximately 2.5\% in magnitude. Correspondingly, the derived nuclear quadrupole moment increases in absolute value by about 2.6\% (from $-0.038177$\,b to $-0.039163$\,b). Similar, albeit slightly smaller, adjustments are observed for LiF and LiCl, with vibrational corrections leading to changes on the order of 1.7--1.9\% in the EFGs and comparable variations in the quadrupole moments.

\begin{figure}
        \centering
        \includegraphics[width=1.0\linewidth]{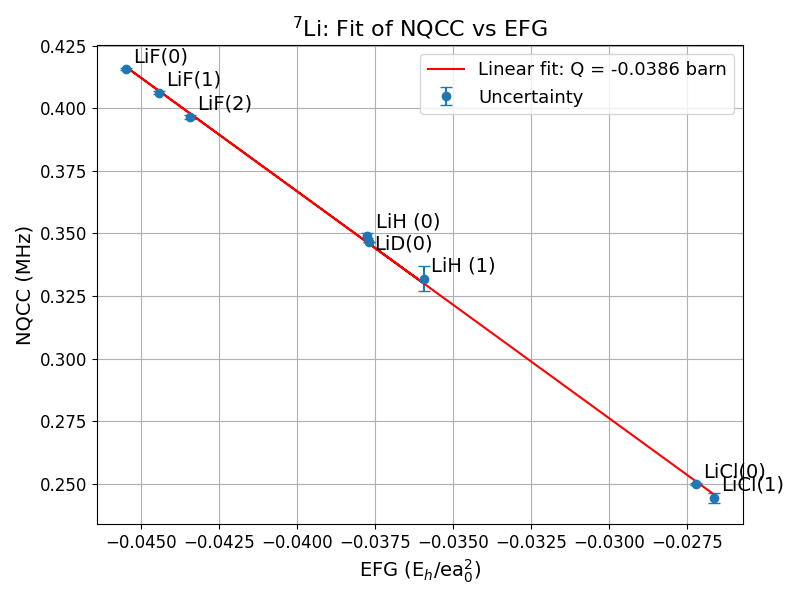}
        \caption{Linear fit of the nuclear quadrupole coupling constant (NQCC) as a function of the electric field gradient (EFG) for the $^{7}$Li nuclei in different compounds. The value of $Q$ is determined from the slope and is expressed in barns, with uncertainties on the NQCC values.}
        \label{plot-linear-li}
    \end{figure}

\begin{table}[ht]
    \centering
    \label{tab:li_efg}
    \sisetup{
        table-number-alignment = center,
        table-format = -1.6
    }
    \begin{tabular}{l l S[table-format=-1.6] S[table-format=-1.6]}
        \toprule
        Molecule & {$\nu,\,J$}  & {EFG ($E_{h}/ea_{0}^{2}$)} & {Q (b)} \\
        \midrule
        $\ce{^{7}Li^{1}H}$  & $\nu=0,\,J=1$ & -0.037682 & -0.039163 \\
        $\ce{^{7}Li^{1}H}$  & $\nu=1,\,J=1$ & -0.035923 & -0.039333 \\
        $\ce{^{7}Li^{2}H}$  & $\nu=0$       & -0.037766 & -0.039329 \\
        $\ce{^{7}Li^{19}F}$ & $\nu=0$       & -0.045475 & -0.038895 \\
        $\ce{^{7}Li^{19}F}$ & $\nu=1$       & -0.044433 & -0.038898 \\
        $\ce{^{7}Li^{19}F}$ & $\nu=2$       & -0.043440 & -0.038847 \\
        $\ce{^{7}Li^{35}Cl}$& $\nu=0$       & -0.027206 & -0.039098 \\
        $\ce{^{7}Li^{35}Cl}$& $\nu=1$       & -0.026649 & -0.039064 \\
        \bottomrule
    \end{tabular}
     \caption{(Ro-)vibrationally averaged EFGs and corresponding NQMs for $\ce{^{7}Li}$.}
\end{table}
In Figure~\ref{plot-linear-li}, a linear regression was performed between the experimental nuclear quadrupole coupling constants and the calculated electric field gradients, as defined in Eq.~(\ref{NQCC}). This best-fit procedure yields a value of $Q(\ce{^7Li}) = -0.0386$\,b, which is slightly lower than the currently recommended molecular value of $-0.0406$ b but identical to the value reported by Guan and co-workers\cite{Guan:2024oyp}, even if obtained by a different route.  

In order to understand this discrepancy, we made a comparison between our  EFG values respect to those reported by  Urban and Sadlej.\cite{URBAN1990157}
At the Hartree--Fock level, the difference is on the order of 0.0002~$E_{h}/ea_{0}^{2}$. The basis set we employed differs from that used by Urban and Sadlej. The basis set used by Urban and Sadlej for LiH is \text{13s8p6d2f}, contracted to \text{9s6p3d1f}, while for LiF and LiCl, they employed the \text{13s8p6d2f} basis set, which, after contraction, becomes \text{8s5p3d1f}. In our case, the same uncontracted basis set was used for all three systems, belonging to the dyall family, which is optimized for relativistic calculations. We selected the \text{dyall.ae4z} basis set as our starting point, which explicitly corresponds to \text{19s10p5d3f}, and we did not contract it. As discussed in the previous section, we added a p- and d-tight exponents, resulting in a \text{19s11p6d3f} basis set. This basis set was chosen as it allows achieving EFG convergence on the order of $10^{-6}$ $E_{h}/ea_{0}^{2}$. with respect to the increase of the basis set exponents. At the CCSD level the discrepancy increases to approximately 0.0005 $E_{h}/ea_{0}^{2}$. The electrons of Li have always been fully correlated in Urban and Sadlej calculations, whereas for F and Cl, the  electrons in the K and L shells have been left uncorrelated. However, there is no information regarding the number of virtual orbitals that were included in the correlation. We may therefore assume that all the virtual orbitals were correlated. In our case, in the CCSD calculation with the \text{19s11p6d3f} basis set, we have correlated all electrons of the atoms and considered all the virtual orbitals. Since our basis set is optimized for properties and because we have correlated a larger number of electrons in the investigated systems (by considering all available virtual orbitals in the active space), we believe that our CCSD value is more accurate than previous results.

However, the most significant deviation arises in the estimation of the contribution from triple excitations, which reaches values as large as 0.002 $E_{h}/ea_{0}^{2}$. for LiF and LiCl. In their paper they used the T(CCSD) method, which is an approximation to reduce the computational cost of including triple excitations. This method is based on the calculation of the contribution of triple excitations to the energy using a fourth-order perturbation theory expression to evaluate the triple excitation.\cite{Urban1985} In contrast, we employ the full CCSDT method without any approximation, except for the limited number of correlated electrons and virtual orbitals, as we pointed out in Table \ref{tab:comparison-li-al}. A further inspection reveals that using a smaller basis set, while retaining all virtual orbitals in the active space, still yields corrections of the same order of magnitude as those reported in Table~\ref{tab:li_efg_contributions}. Therefore, we argue that their estimate of the contribution from triple excitations is too high by approximately one order of magnitude compared to our fully iterative treatment.

We shall conclude by noting that the recent results provided by Guan and co-workers, based on high-precision measurements of hyperfine splittings in the $2^{3}P_J$ state of ${}^{7}$Li$^{+}$, \cite{Guan:2024oyp} provide a value of $Q(\ce{^7Li})=$-0.0386(5) b, which is in perfect agreement with our value (without including QED corrections), and therefore we believe that the currently accepted value of the quadrupole moment should be revisited. 

\newpage
\subsection{Nuclear Quadrupole Moment of  $^{27}$Al from molecular data}

The reference value of the nuclear quadrupole moment of \(^{27}\mathrm{Al}\) is \(0.1466 \pm 0.0010\)b, as reported by Kellö and co-workers,\cite{Kellö1991} based on atomic and molecular calculations. This was later confirmed by Pernpointner and Visscher via a four-component relativistic CCSD(T) finite-field calculation, yielding \(0.1460 \pm 0.0004\)b.\cite{Pernpointner2001_JCP} Both studies considered the AlY (Y = F, Cl) systems, and Pernpointner and Visscher additionally included AlBr to assess relativistic contributions to \(Q(^{27}\mathrm{Al})\). Brown and Wasylishen\cite{Brown2013_JMS} reported a larger value, \(Q(^{27}\mathrm{Al}) = 0.149 \pm 0.002\)b (the uncertainty stemming mainly from experimental error), based on EFG calculations in AlH and AlD.\cite{9d88811980d04af9a5ca70a30b620871} More recently, Aerts and Brown\cite{Aerts2019_JCP} proposed a revised value \(Q(^{27}\mathrm{Al}) = 0.1482 \pm 0.0005\)b, neglecting relativistic effects but including vibrational averaging.

Given these recent results, which differ from previous findings, the necessity of further investigating the quadrupole moment of $^{27}\text{Al}$ becomes evident. Moreover, as previously shown, triple excitations still have a non-negligible impact on the value of \( Q(\ce{^{27}Al}) \). We will consider the systems AlY (Y=H, F, Cl, Br) and employ the methodology previously discussed, where we will compute the reference value at the CCSD level with a sufficiently large active space and an extended basis set, before introducing triple and quadruple corrections. Finally, we will compute the vibrationally-averaged electric field gradient.

Table \ref{tab:al_efg_contributions} details the stepwise contributions to the electric-field gradient (EFG) at the aluminum nucleus and the corresponding derived values of the nuclear quadrupole moment (NQM) for the series AlY (Y = H, F, Cl, Br). The starting Hartree–Fock values show significant deviation from the final results, highlighting the critical role of correlation effects. The inclusion of CCSD-level correlation already improves the EFG and $Q$ values significantly.

Triple excitations introduce non-negligible corrections to the EFG (ranging from approximately 0.006 to 0.009 $E_{h}/ea_{0}^{2}$), which systematically affect the computed quadrupole moments by about 1 mb, underlining their importance even at this level of theory. Quadruple corrections are smaller, but still contribute meaningfully in the lighter systems. 

Overall, the data confirm that an accurate determination of the NQM requires the inclusion of triple excitations and vibrational effects, especially for the light ligands.

\begin{sidewaystable}[h!]
    \centering
    \begin{tabular}{lccccccccccc}
        \toprule
        & \multicolumn{2}{c}{AlH} & \multicolumn{2}{c}{AlD} & \multicolumn{2}{c}{AlF} & \multicolumn{2}{c}{AlCl} & \multicolumn{2}{c}{AlBr} \\
        \cmidrule(lr){2-3} \cmidrule(lr){4-5} \cmidrule(lr){6-7} \cmidrule(lr){8-9} \cmidrule(lr){10-11}
        & EFG & $Q$ & EFG & $Q$ & EFG & $Q$ & EFG & $Q$ & EFG & $Q$ \\
        \midrule
        HF & -1.472185 & 0.140450 & -1.472185 & 0.140450 & -1.192554 & 0.134721 & -0.981921 & 0.131798 & -0.906967 & 0.131418 \\
        CCSD  & -1.4208798 & 0.145542 & -1.4208798 & 0.145542 & -1.106585 & 0.145188 & -0.897732 & 0.144158 & -0.828187 & 0.143919 \\
        + correction for full triples & 0.008609 & 0.001154 & 0.008609 & 0.001154 & 0.008726 & 0.001154 & 0.006460 & 0.001045 & 0.005803 & 0.001015 \\
        + corrections for full quadruples & 0.000405 & 0.000042 & 0.000405 & 0.000042 & 0.000022 & 0.000003 & 0.000024 & 0.000004 & 0.000005 & 0.000000 \\
        + vibrational corrections & 0.017228 & 0.001809 & 0.012242 & 0.000947 & -0.000217 & -0.000029 & 0.003110 & 0.000509 & 0.0148850 & 0.002672 \\
        final value & -1.394638 & 0.148547 & -1.399624 & 0.147685 & -1.098055 & 0.146315 & -0.888138 & 0.145716 & -0.807495 & 0.147607 \\
        \bottomrule
    \end{tabular}
    \caption{Individual contributions to the EFG (in $E_{h}/ea_{0}^{2}$) at the Al position for the systems AlY (Y = H, D, F, Cl, Br) and the derived values for the $^{27}$Al quadrupole moment ($Q$, in mb).}
    \label{tab:al_efg_contributions}
\end{sidewaystable}

\begin{figure}
    \centering
    \includegraphics[width=1.0\linewidth]{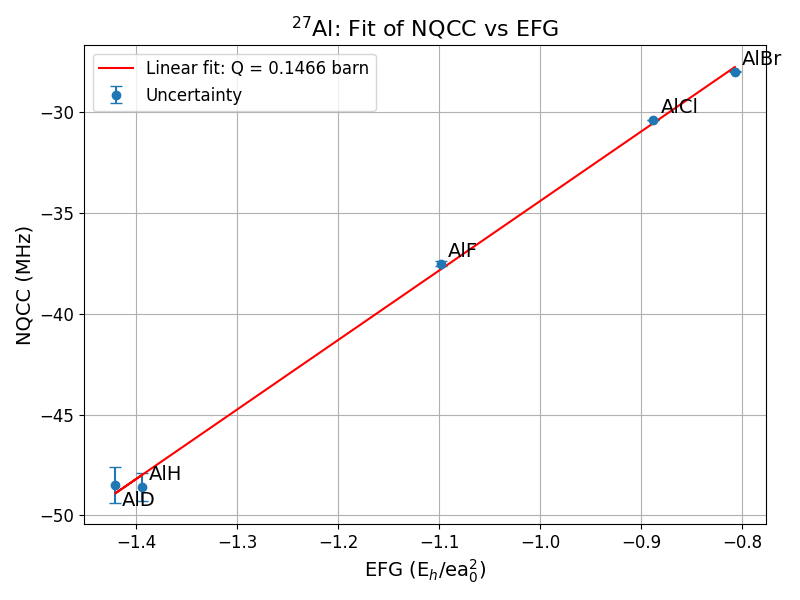}
    \caption{Linear fit of the nuclear quadrupole coupling constant (NQCC) as a function of the electric field gradient (EFG) for the $^{27}$Al nuclei in different compounds. The value of $Q$ is determined from the slope and is expressed in barns, with uncertainties on the NQCC values.
}
    \label{linear-fit-al}
\end{figure}

In Figure \ref{linear-fit-al} we made a linear regression of the experimental nuclear quadrupole coupling constants versus the calculated electric field gradients (according to Eq. (\ref{NQCC})) which yielded a best fit $Q(\ce{^{27}Al})=0.1466$~b, which shows perfect agreement with previous studies.\cite{Kello1999_CPL,Pernpointner2001_JCP} The slightly higher values obtained by Brown and Wasylishen\cite{Brown2013_JMS} can be essentially attributed to the inaccuracy of the experimental NQCCs, which, in the AlY series (Y = H, F, Cl, Br), are affected by the largest uncertainties, as shown in Table \ref{tab:aluminum_spectroscopy}. In our case, since we include a larger number of systems—many of which have more precisely determined NQCCs—and take into account the experimental uncertainties through linear regression, we obtain a quadrupole moment that is consistent with the accepted reference value. More recent studies have proposed a slightly higher value, such as $Q(^{27}\mathrm{Al}) = 0.1482$b  by Aerts and Brown,\cite{Aerts2019_JCP} but it is important to note that their approach neglected relativistic effects and do not present a basis set convergence test, which we claim is quite important, as clearly demonstrated by van Stralen and Visscher.\cite{VANSTRALEN10072003} In contrast, our methodology explicitly accounts for these corrections.

\section{\label{sec:level6}Conclusions}
In this work, we implemented CCSDT and CCSDTQ expectation values within the DIRAC quantum chemistry program.\cite{DIRAC2020} We demonstrated how the implementation can be efficiently carried out using the \text{tenpi} toolchain\cite{brandejs2024generatingcoupledclustercode} for the generation of $\Lambda$ equations and density matrices. To validate the CCSDT and CCSDTQ expectation value codes, we considered the LiH system and computed its dipole moment at the non-relativistic level from CCSD up to CCSDTQ, finding excellent agreement with the well-established MRCC quantum chemistry program.\cite{kallay2020mrcc}

 A more significant application was the accurate calculation of the quadrupole moment for the nuclei $^7\text{Li}$ and $^{27}\text{Al}$ from molecular data. We have performed a detailed analysis of the electric field gradients at the lithium and aluminum nuclei for the series of LiX (X = H, F, Cl) and AlY (Y = H, F, Cl, Br) compounds. By employing CCSD calculations with an extended basis set, further corrected through the inclusion of full triple and quadruple excitations and vibrational averaging, we derived the corresponding nuclear quadrupole moments. For the lithium atom, we found a value of $Q(^{7}\mathrm{Li}) = -0.0387$\,b, slightly lower than the currently accepted reference value, but in perfect agreement with Guan and co-workers.\cite{Guan:2024oyp} For aluminum, our final data yield a value of $Q(^{27}\mathrm{Al}) = 0.1466$\,b, in excellent agreement with the molecular values reported by Kellö \textit{et al.}\,\cite{Kellö1991} and confirmed by Pernpointner and Visscher.\cite{Pernpointner2001_JCP}

At present, our implementation faces three main bottlenecks that limit its scalability and overall efficiency: (i) the restriction to a single node, and since we are not exploiting (ii) point-group symmetry and (iii) index anti-symmetry of tensors. These limitations hinder both performance and memory efficiency, especially when tackling larger molecular systems or higher-order correlation methods. To address these issues, we are actively working on several fronts. First, we are transitioning towards a distributed-memory, multi-node implementation based on the Cyclops Tensor Framework.\cite{solomonik2014massively} Second, we are implementing point-group symmetry to reduce the number of independent amplitudes and integrals, which will also allow us to exploit the permutation antisymmetry of tensors, thus accelerating calculations and reducing memory requirements. These developments are expected to significantly improve the computational efficiency and make high-accuracy methods feasible for larger and more complex molecular systems. However, these technical developments go beyond the scope of the present work.

The implementation of CCSDT and CCSDTQ expectation values in DIRAC represents a further step forward in the HAMP-vQED project.\cite{Saue2024} Combined with the inclusion of QED effects, on which we are currently working,\cite{Sunaga2022_JCP,Salman2023_APS,ferenc2025gaussianbasissetapproach} this will enable the highly accurate calculation of molecular properties.

\begin{acknowledgments}
We dedicate this paper to the memory of John F. Stanton.

This project was funded by the European Research Council (ERC) under the European Union’s Horizon 2020 research and innovation program (grant agreement ID:101019907). This work was performed using HPC resources from CALMIP (Calcul en Midi-Pyrenées; Grant 2024-P13154 and 2024-M24070).
\end{acknowledgments}

\section*{Author declarations}
\subsection*{Conflict of Interest}
The authors declare that they have no conflict of interest. 
\section*{Data Availability Statement}
The data that support the findings of this study will be made openly available in ZENODO (after acceptance of the manuscript).\cite{ReplicationData}
\bibliography{manuscript}
\newpage
\section{TOC}
\begin{figure}[htbp]
  \centering
  \begin{tikzpicture}[node distance=1cm and 1.0cm, auto, >=Latex] 
    \tikzset{
      gradientBox/.style={
        rectangle,
        draw=black!50,
        top color=blue!10,
        bottom color=blue!30,
        shading=axis,
        shading angle=90,
        minimum width=7cm,
        minimum height=1cm,
        rounded corners=4pt,
        drop shadow={shadow xshift=0.5ex,shadow yshift=-0.5ex,opacity=0.3}
      },
      arrow/.style={->, thick, black!70}
    }

    \node[gradientBox] (dirac) {\textbf{DIRAC}: ExaCorr module implementation};
    \node[gradientBox, below=of dirac] (t) {Solve amplitude equations};
    \coordinate (mid) at ($(dirac)!0.5!(t)$);
    \node[gradientBox, right=5.5cm of mid, minimum width=3cm] (tenpi) {\textbf{tenpi.py}}; 

    \draw[arrow] (dirac) -- (t);
    \draw[arrow] (tenpi) -- node[above]{\text{t.F90}, $\Lambda$\text{.F90}, $\gamma^{CC}$\text{.F90}} (mid); 

    \node[gradientBox, below=of t] (lambda) {Solve $\Lambda$-equations using converged t amplitudes};
    \draw[arrow] (t) -- (lambda);

    \node[gradientBox, below=of lambda] (density) {Compute $\tilde{\gamma}^{CC}$ using converged t and $\overline{t}$ amplitudes};
    \draw[arrow] (lambda) -- (density);

    \node[gradientBox, below=of density] (expect) {Expectation value: $\frac{dL_{CC}}{d\varepsilon_{A}}\Big|_{\bm{\varepsilon}=0} = \sum_{pq} h_{pq;A}\,\tilde{\gamma}^{CC}_{pq}$};
    \draw[arrow] (density) -- (expect);
  \end{tikzpicture}
  \label{fig:exacorr_workflow}
\end{figure}

\end{document}